\documentclass[aps,pra,amsmath,amssymb]{revtex4-1}
\usepackage{pgf,tikz}
\usepackage{tensor}
\usepackage{graphics}
\usepackage{graphicx}
\usepackage{bm}
\usepackage{amsmath}
\usepackage{amsfonts}
\usepackage{amssymb}
\usepackage{latexsym}
\usepackage{color}
\usepackage{bbold}
\usepackage{braket,dsfont,bm}
\usepackage{pgfplots}
\usepackage{blochsphere}
\usepackage{mathrsfs}
\usetikzlibrary{arrows}
\usepackage[sc]{mathpazo}

\begin{document}

\title{Journeys from Quantum Optics to Quantum Technology}

\author{Stephen M. Barnett$^1$}
\author{Almut Beige$^2$}
\author{Artur Ekert$^3$}
\author{Barry M.\ Garraway$^4$}
\author{Christoph H. Keitel$^{5}$}
\author{Viv Kendon$^6$}
\author{Manfred Lein$^7$}
\author{Gerard J. Milburn$^{8}$}
\author{H\'ector M. Moya-Cessa$^{9}$}
\author{Mio Murao$^{10}$}
\author{Jiannis K. Pachos$^{2}$}
\author{G. Massimo  Palma$^{11}$}
\author{Emmanuel Paspalakis$^{12}$}
\author{Simon J. D. Phoenix$^{13}$}
\author{Bernard Piraux$^{14}$}
\author{Martin B. Plenio$^{15}$}
\author{Barry C.\ Sanders$^{16,17,18,19}$}
\author{Jason Twamley$^{20}$}
\author{A. Vidiella-Barranco$^{21}$}
\author{M. S. Kim$^{22}$}

\affiliation{$^1$ School of Physics and Astronomy, University of Glasgow, Glasgow G12 8QQ, UK}
\affiliation{$^2$ The School of Physics and Astronomy, University of Leeds, Leeds LS2 9JT, UK}
\affiliation{$^3$ Mathematical Institute, University of Oxford \& Centre for Quantum Technologies, National University of Singapore.} 
\affiliation{$^4$ Department of Physics and Astronomy, University of Sussex, Falmer, Brighton, BN1 9QH, UK}
\affiliation{$^5$  Max-Planck-Institut f\"{u}r Kernphysik, Saupfercheckweg 1, 69117 Heidelberg, Germany}
\affiliation{$^6$ Department of Physics, Durham University, South Road, Durham DH1 3LE, UK}
\affiliation{$^7$ Institute for Theoretical Physics, Leibniz Universit\"at Hannover, 30167 Hannover, Germany}
\affiliation{$^{8}$Centre for Engineered Quantum Systems, School of Mathematics and Physics, The University of Queensland, Australia, 4072. }
\affiliation{$^{9}$ Instituto Nacional de Astrof\'{\i}sica, \'{O}ptica y Electr\'{o}nica, Calle Luis Enrique Erro No.\ 1, Sta. Ma. Tonantzintla, Pue.\ CP 72840, Mexico}
\affiliation{$^{10}$ Department of Physics, Graduate School of Science, The University of Tokyo, Japan}
\affiliation{$^{11}$ NEST, Istituto Nanoscienze-CNR and Dipartimento di Fisica e Chimica, Universit$\grave{a}$  degli Studi di Palermo, via Archirafi 36, I-90123 Palermo, Italy}
\affiliation{$^{12}$ Materials Science Department, School of Natural Sciences, University of Patras, Patras 265 04, Greece}
\affiliation{$^{13}$ Department of Applied Mathematics and Sciences, Khalifa University, P.O. Box: 127788, Abu Dhabi, UAE}
\affiliation{$^{14}$ Laboratoire de Physique Atomique, Moleculaire et Optique (unite PAMO), Universite Catholique de Louvain,  B-1348 Louvain-la-Neuve, Belgium}
\affiliation{$^{15}$ Institute of Theoretical Physics and Center for Integrated Quantum Science and Technology (IQST), Albert-Einstein-Allee 11, Universit\"at Ulm, 89069 Ulm, Germany}
\affiliation{$^{16}$ Institute for Quantum Science and Technology, University of Calgary, Alberta T2N 1N4, Canada}
\affiliation{$^{17}$ Hefei National Laboratory for Physical Sciences at Microscale,
	University of Science and Technology of China, Hefei, Anhui 230026, China}
\affiliation{$^{18}$ Shanghai Branch, CAS Center for Excellence and Synergetic Innovation Center
		in Quantum Information and Quantum Physics,\\
	University of Science and Technology of China, Shanghai 201315, China}
\affiliation{$^{19}$ Program in Quantum Information Science, Canadian Institute for Advanced Research, Toronto, Ontario M5G 1Z8, Canada}
\affiliation{$^{20}$ Centre for Engineered Quantum Systems, Department of Physics and Astronomy, Macquarie University, Sydney, NSW 2109, Australia}
\affiliation{$^{21}$ QOLS, Blackett Laboratory, Imperial College London, SW7 2AZ, UK}
\affiliation{$^{22}2$ Instituto de F\'\i sica ``Gleb Wataghin'', Universidade Estadual de Campinas - UNICAMP, 13083-859, Campinas, SP, Brazil}

\date{\today}

\begin{abstract}
Sir Peter Knight is a pioneer in quantum optics which has now grown to an important branch of modern physics to study the foundations and applications of quantum physics. He is leading an effort to develop new technologies from quantum mechanics. In this collection of essays, we recall the time we were working with him as a postdoc or a PhD student and look at how the time with him has influenced our research.

\end{abstract}

\maketitle

\section{Introduction}
Quantum technology is rapidly becoming one of the most viable new technologies, showing a great potential to provide a major breakthrough in the 21st century industry \cite{economist}. There are several reasons for this confidence.  Most of the important components have been developed for quantum technology and proofs of principles have been demonstrated. We need to combine all the components at a large scale, which is challenging but there is no fundamental reason why this is not possible. In the past decades, quantum researchers have identified problems, many of which have been solved. The research community has grown enormously to cover a large interdisciplinary research areas from physics and biology to electrical, computational and civil engineering. The growing size of a skilled work force is taking the research from strength to strength. There are also important reasons which make the advent of the new technology necessary. With the limited amount of resources on Earth, it is impossible to project permanent growth in the global economy, unless we are able to use the resources efficiently. To this end, it is obvious that we should be able to control the properties of materials to the smallest scale: atomic level, where the law of nature is governed by quantum mechanics. Secondly, many core technologies are already facing the limit for further developments.  Delving further into the same ideas only yields improvements in decreasing increments and will not bring about new technologies which can overcome the shortcomings. While the existing technology is mainly based on classical physics and conventional statistics, quantum physics generalise them to a much larger mathematical space, opening a possibility to utilise nature more efficiently. 

In order to realise the potential that quantum technology offers, a coherent and strategic action is required. Recognising this, a few government-level initiatives have been developed. Sir Peter Knight is leading the UK effort to develop quantum technology. As a pioneer of quantum optics, he has made a considerable contribution to further our understanding of quantum mechanics and nature of non-classicalities. Recently, his efforts have been directed at applying such knowledge to control quantum-mechanical interactions to identify the advantages offered by quantum mechanics and utilise them to revolutionise technology in various corners of industry including sensing, computing, imaging and communication.  In the early days of his career, Peter Knight worked intensively on high harmonic generation and interaction of atoms and intense laser fields like some other quantum optics researchers. Peter Knight is not only a pioneering researcher but also an inspiring teacher. In this collection of essays, we recall the times we were working with him as a postdoc or a PhD student and present how the time with him has influenced our research.

\section{Thermal states are mixed \dots aren't they? --  Barnett}
\label{BARNETT}

In the early spring of 1984, Peter Knight was away for a week or two and I was shown, by Claire Gilson, a paper on 
thermofield dynamics \cite{Takahashi75}, which contained a remarkable idea.  This was the demonstration that
it is possible to express thermal averages in quantum field theory as vacuum expectation values for a ``temperature-dependent 
vacuum" state in a doubled state-space.  The temperature-dependent  vacuum state is related to the true vacuum state, moreover,
by a unitary transformation; a transformation that I recognised as being a Bogoliubov transformation familiar from 
squeezed states \cite{PLKRodney} which were, at the time, becoming a hot topic in quantum optics.  The idea that
we could represent a thermal state of a radiation field as a \emph{pure} state was something of a revelation and I
have to say that Peter took some convincing of it, but we were intrigued by the idea and proceeded to apply it to
a range of phenomena in quantum optics \cite{thermoQO}.  I present here a brief introduction to the thermofield
formalism and some of the ideas that it led on to including what are now known as purification and the quantum
mutual information \cite{QIbook}.

The expectation value of an observable $\hat{A}$ for a quantum system in a thermal state is
\begin{equation}
\langle \hat{A}\rangle = {\rm Tr}\left(\hat{A}\hat{\rho}_{\rm th}\right) = {\rm Tr}\left(\hat{A} \text{e}^{-\beta \hat{H}}\right)/Z(\beta) ,
\end{equation}
where $\hat{\rho}_{\rm th} = \text{e}^{-\beta \hat{H}}/Z(\beta)$ is the density operator for the thermal state, $\hat{H}$ is the Hamiltonian,
$\beta = (k_{\rm B} T)^{-1}$ is the inverse temperature and $Z(\beta)$ is the partition function.  The thermofield 
representation of this average is the \emph{pure-state} expectation value
\begin{equation}
\langle \hat{A}\rangle = \langle 0(\beta)|\hat{A}|0(\beta)\rangle ,
\end{equation}
where $|0(\beta)\rangle$ is the temperature-dependent vacuum state.  We can try to construct this state as a superposition
of energy eigenstates:
\begin{equation}
|0(\beta)\rangle = \sum_n |E_n\rangle f_n(\beta) 
\end{equation}
which leads to the requirement that $f^*_n(\beta)f_m(\beta) = Z^{-1}(\beta)\text{e}^{-\beta E_n}\delta_{nm}$, so that the $f_n$
need to be orthogonal vectors.  The natural, and indeed easiest, way to achieve this is to introduce a second ``ficticious"
state space, which we denote with a tilde, so that our temperature-dependent vacuum state becomes
\begin{equation}
|0(\beta)\rangle = Z^{-1/2}(\beta) \sum_n \text{e}^{-\beta E_n/2}|E_n, \tilde{E}_n\rangle ,
\end{equation}
which we recognise as a pure state of two identical systems which are entangled in their energy.

Can we play the same game and represent \emph{any} mixed state of a quantum system as a pure state?  The answer,
of course, is yes all we need is an entangled pure state in a doubled state-space with the Schmidt basis for the pure 
state corresponding to the basis in which the mixed state is diagonal.  The proof of validity of this simple procedure,
which later became known as purification, is in \cite{Liouville}.

Much of the power of the thermofield technique follows from the observation that for bosons (and indeed for fermions)
the temperature-dependent vacuum state is simply related to the true vacuum by a simple linear transformation.  For
photons, for example, we have
\begin{equation}
|0(\beta)\rangle = \hat{T}(\theta)|0,\tilde{0}\rangle = \exp\left[\theta(\beta)\left(\hat{a}^\dagger\hat{\tilde{a}}^\dagger - \hat{\tilde{a}}\hat{a}\right)\right]|0,\tilde{0}\rangle ,
\end{equation}
where $\tanh\theta(\beta) = \text{e}^{-\beta\hbar\omega/2}$ and $\omega$ is the optical angular frequency.  The consequence of 
this is that we can convert expectation values into vacuum expectation values by performing the unitary transformations
on the operators:
\begin{eqnarray}
\hat{T}^\dagger(\theta)\hat{a}\hat{T}(\theta) &=& \hat{a}\cosh\theta + \hat{\tilde{a}}^\dagger\sinh\theta \nonumber \\
\hat{T}^\dagger(\theta)\hat{a}^\dagger \hat{T}(\theta) &=& \hat{a}^\dagger\cosh\theta + \hat{\tilde{a}}\sinh\theta ,
\end{eqnarray}
so that, for example,
\begin{eqnarray}
\langle \hat{a}^\dagger \hat{a}\rangle &=& \langle 0(\beta)|\hat{A}|0(\beta)\rangle  \nonumber \\
&=& \langle 0,\tilde{0}|(\hat{a}^\dagger\cosh\theta + \hat{\tilde{a}}\sinh\theta)(\hat{a}\cosh\theta + \hat{\tilde{a}}^\dagger\sinh\theta)|0,\tilde{0}\rangle
\nonumber \\
&=& \sinh^2\theta(\beta) = \frac{1}{\text{e}^{\beta\omega}-1} .
\end{eqnarray}
This technique becomes very useful in thermal field theory \cite{UmezawaQFT,Matsumoto} and in many-body physics 
\cite{BJDJJ}.

What caught our attention, however, was that the state $|0(\beta)\rangle$ is formally a two-mode squeezed vacuum 
state \cite{Gerard} and that it becomes this state if we replace the ficticious mode by a real one.  It followed that these
real two-mode states have single-mode thermal properties.  They were also pure states, however, so the mixed-state
behaviour of the single modes must be a consequence of the entanglement between the two modes \cite{SqCorr,Ekert}.
This idea was also very much in our minds at that time because of Peter's work with Simon Phoenix on entropy in
the Jaynes-Cummings model and its relatives \cite{Annals}.  From this we knew, in particular, that for a pure state 
of two entangled quantum systems, the entropies for each have precisely the same value, however, complicated
the form of the state or the dynamics that generated it.

The fact that we had, in the two-mode squeezed vacuum, a pure state with each individual mode in a thermal state 
\cite{thermoQO} led to a simple chain of reasoning: (i) Thermal mixed states are those with the greatest entropy for
a given mean energy, which also means that they are maximally uncertain.  (ii) Therefore, as the two mode state is 
pure, this must mean that these states have, in sense, the greatest amount of information hidden in the correlations
between the modes, and therefore (iii) they seem to be the most strongly entangled states for a given energy.  This
idea was formalised by introducing the ``index of correlation" \cite{Simon1,Simon2}
\begin{equation}
I_c = S_a + S_b - S ,
\end{equation}
where $S_{a(b)}$ is the von Neumann entropy for the $a(b)$ mode and $S$ is that for the combined two-mode state.
This quantity, which is now well-known in quantum information theory as an entanglement measure \cite{QIbook},
is maximised (for a given mean photon number) for the two-mode squeezed vacuum state.

Looking back some thirty or so years I pause to wonder if we had any inkling that these ideas would play such an 
important part in the development of the field \emph{quantum information} that was about to begin.  We knew certainly
that quantum superpositions meant that there were many more ways in which information could be encoded on a
quantum system than on a classical one and, moreover, that communication laboratories were starting to become
very interested in non-classical states of light.  But the honest answer is ``no" we did not - we were just playing.

\section{Realising quantum technology with open quantum systems -- Beige}
In 1996, Peter Knight lectured at a Quantum Optics winter school at the Abdus Salam International Centre for Theoretical Physics in Trieste. Back then, my PhD supervisor Gerhard Hegerfeldt and I were working on the quantum Zeno effect \cite{Hegerfeldt}. Many people warned me that working on the foundations of quantum physics was not a promising way of starting a research career but Peter Knight recognised the importance of what we were doing. When I met him in Trieste, we had a detailed scientific discussion. Later on, after completing my PhD, Peter invited me to join his group in London and to work with him on quantum computing with quantum optical systems. I happily moved from a relatively small town in Germany to London to become part of a very likely and inspiring community of theoreticians and experimentalists. At the time, research into quantum computing was still in its infancy and comprised only a relatively small scientific community. 

\subsection{Quantum Computing using dissipation}

Of course, Peter had the right intuition. When I first arrived at Imperial College London, I helped to analyse a probabilistic entangling scheme for two atomic qubits inside an optical cavity \cite{Martin}. Our idea was to initially prepare the atoms in a product state with one atom excited and one atom in its ground state. Under the condition of no photon emission through the cavity mirrors, we could show that the atoms ended up in a maximally entangled state. Later, I realised that the quantum Zeno effect could be used to push the success rate of entangling schemes in open quantum systems up to $100\,\%$.
After analysing concrete entangling scheme \cite{NJP}, we used this idea to design deterministic quantum gate operations on atomic qubits inside an optical resonator \cite{letter}. 

The ground states of the atomic qubits were part of a so-called decoherence-free subspace. Continuously measuring whether or not the system remained in its coherence-free state, thereby allowing us to take advantages of the quantum Zeno effect, resulted in the implementation of an effective Hamiltonian of the form
\begin{eqnarray}
\hat{H}_{\rm eff} &=& I \!\! P_{\rm DS} \, \hat{H}_{\rm sys} \, I \!\! P_{\rm DS} \, .
\end{eqnarray}
Here $I \!\! P_{\rm DS}$ denotes the projector onto the decoherence-free subspace and where $\hat{H}$ denotes the actual Hamiltonian of the open quantum system. Even when $\hat{H}_{\rm sys}$ does not contain any direct interactions between qubits, $\hat{H}_{\rm eff} $ might. 

Showing that dissipation does not limit but can actually enhance the performance of quantum computing devices made us excited and optimistic. Unfortunately, the prize we had to pay was time. Using the quantum Zeno effect might protect quantum gate operations against one type of dissipation (like cavity decay) but the corresponding increase in operation time invites other problems. For example, in Ref.~\cite{letter}, it increased the overall probability for the emission of photons from excited atomic states which is accompanied by the loss of information. The implementation of the quantum gate operation proposed in Ref.~\cite{letter} still required a relatively strong atom-cavity coupling constant compared to the relevant spontaneous decay rates. To overcome this problem, we later designed alternative schemes with an even stronger protection while taking advantage of dissipation. These schemes were based on dissipation-assisted adiabatic passages \cite{Rempe} or on macroscopic light and dark periods \cite{Metz} or simply cooled atoms into entangled states \cite{Spiller}. 

Open quantum systems can be used to implement a wide range of computational tasks. For example, in Ref.~\cite{Lim} we showed that it is in principle possible to implement deterministic entangling gate operations between distant atomic qubits through photon measurements. Up to then it was believed that the success rate of quantum gate operations for distributed quantum computing could not exceed $50 \, \%$. Our way around this limitation was to explore a mutually unbiased basis and to repeat operations until success. More recently we proposed to use an optical cavity inside an instantaneous quantum feedback loop for measurements of the phase shift between two pathways of light \cite{Lewis}. Although our scheme only uses single-mode coherent states and no entanglement is present, our quantum metrology scheme manages to exceed the standard quantum limit. The origin of this enhancement is quantum feedback-induced non-linear dynamics with a wide range of potential applications in quantum technology.

\subsection{Modelling open quantum systems}

The above results would not have been possible without accompanying more fundamental research. For example, the idea of using dissipation for quantum computing was inspired by our previous work on the quantum Zeno effect \cite{Hegerfeldt}. The idea of using interference effects for distributed quantum computing was inspired by studying interference effects in a related two-atom double slit experiment \cite{Schoen}. There are still many open questions in quantum optics, like ``What exactly are photons?" \cite{Bennett}. Other open questions are, ``Is there a Hamiltonian to describe the propagation of light through a beamsplitter?'' \cite{Nick} and "How can we model the light scattering from two-sided optical cavities?" \cite{Tom}. Answering these questions will help us to find even more efficient solutions for applications in quantum technology and to design and analyse novel quantum systems, like coherent cavity networks with complete connectivity \cite{Elica}.

\section{Born's rule, morris dancing and the importance of English ale -- Ekert}
We were three pints each into one of those long evenings at The Eagle and Child (aka The Bird and Baby) when Peter switched rapidly from drawing distorted contours of the Wigner function (our two-mode squeezed states paper had been accepted but required some last minute amendments) into the quirks of the British parliamentary system. Not a big surprise; by then I had already estimated that it took Peter, on average, $\pi$ pints of ale to reach this stage. Truth be told, I eagerly waited for these digressions. Peter's encyclopaedic knowledge combined with his gift of storytelling made his anecdotes anything but boring. Moreover, he truly believed that his supervision duties extended well beyond the Jaynes-Cummings model and covered, among other things, the relevance of the distance between the front benches in the House of Commons (precisely the length of two drawn swords), the intricacies of Morris Dancing (beardy men in white stockings prancing around to accordion music waving hankies and hitting each other with sticks), tips on how to win at Mornington Crescent (I am sorry, Pete, I still haven't a clue) and many other delightful British idiosyncrasies.

By the fourth round we were back to the world of two-mode squeezed states and spooky actions at a distance. To be sure, the Inklings we were not, but our own quantum fantasy world (or many worlds in my case) could beat Mordor and Narnia hands down. By the end of the fifth round we were taking apart the Born rule. `Just think about it. Why do we \emph{square} the amplitudes, why not raising them to power $42$ or some other meaningful number?\,`, Peter was clearly warming up for his pep talk and I had no good answers. `Should the rule be postulated or rather derived from the formalism of quantum theory?\;`, he continued. Here I was pretty sure it should be derived and mumbled something about Gleason's theorem~\cite{gle57} (which says roughly that the Born rule is the only consistent probability rule you can conceivably have that depends on the state vector alone) but we both quickly agreed that it offered very little in terms of physical insights. Neither of us liked Gleason's theorem, but we could hardly agree on the alternatives, which smoothly propelled our discussion to another round, at which point Peter mentioned that even Max Born didn't quite get Born's rule right, at least not on his first approach. This I found puzzling. How could one of the pioneers of quantum theory got it wrong? I felt some affinity for Born, after all we were born in the same town, just separated by eighty years of its turbulent history, so I decided to learn a bit more about the origins of the rule. 

I cannot recall the end of that particular evening (for a good reason) but I do remember that the following day I visited the Radcliffe Science Library (this is how we did re-search in the pre-Google era). Indeed, in the original paper proposing the probability interpretation of the state vector (wavefunction)~\cite{born26} Born wrote:  
\begin{quote}
...If one translates this result into terms of particles only one interpretation is possible. $\Theta_{\eta,\tau,m}(\alpha,\beta,\gamma)$ [the wavefunction for the particular problem he is considering] gives the probability$^*$ for the electron arriving from the $z$ direction to be thrown out into the direction designated by the angles
$\alpha,\beta,\gamma$... .

$^*$ Addition in proof: More careful considerations show that the probability is proportional to the square of the quantity $ \Theta_{\eta,\tau,m}(\alpha,\beta,\gamma)$.
\end{quote}

\noindent  There was something encouraging and comforting in reading all this. Suddenly Max Born stepped down from his pedestal and became a mere mortal, a human being to whom I could relate. His paper was first rejected by \emph{die Naturwissenshaften} and when it was later accepted by \emph{Zeitschrift f{\"u}r Physik}, he didn't get the Born rule right. I dropped my little historical investigation soon after, but the question `why squaring the amplitudes ?'  has lingered somewhere in the back of my mind ever since. Indeed, why not the amplitude to the fourth, or whatever? 

Suppose probabilities are given by the absolute values of amplitudes raised to power $p$. The admissible physical evolutions must preserve the normalisation of probability. Mathematically speaking, they must be isometries of $p$-norms. Recall that the $p$-norm of vector $v$, with components $v_1, v_2,..., v_n$, is defined as
$
\sqrt[\leftroot{1}\uproot{3}p]{|v_1|^p+|v_2|^p+... |v_n|^p}.
$
It is clear that any permutation of vector components $v_k$ and multiplying them by phase factors (unit complex numbers) will leave any $p$-norm unchanged. Are there any other isometries of $p$-norms? It turns out that these complex permutations are the only isometries, except one special case, namely $p=2$. In this particular case the isometries are unitary operations, which form a continuous group. In all other cases we are restricted to discrete permutations ~\cite{LiSo}. We do not have to go into details of the proof for we can \emph{see} this result. Let us draw unit spheres in different $p$ norms, e.g. for $p=1,2,42$, and $\infty$. 

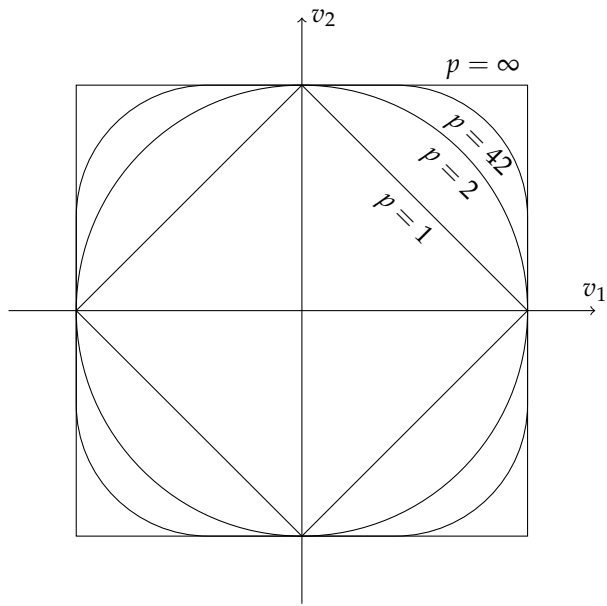
\begin{center}
\begin{tikzpicture} [scale=3]
\draw [->](-1.3,0) -- (1.3,0) node[above]{$v_1$};
\draw [->](0,-1.3) -- (0,1.3) node[right]{$v_2$};
\draw (0,0) circle (1);
\draw (-1,-1) rectangle (1,1);
\draw (1,0) -- (0,1) -- (-1,0) -- (0,-1) -- cycle;
\draw [rounded corners=50pt] (-1,-1) rectangle (1,1);
\node[label={[rotate=-45]below: ${\tiny{p=1}}$}] at (0.52,0.52) {};
\node[label={[rotate=-45]below: ${\tiny p=2}$}] at (0.72,0.72) {};
\node[label={[rotate=-45]below: ${\tiny p=42}$}] at (0.85,0.85) {};
\node[] at (0.8,1.07) {$p=\infty$};
\end{tikzpicture} 
\end{center}

\noindent The image of the unit sphere must be preserved under probability preserving operations. As we can see the $2$-norm is special because of its rotational invariance -- the probability measure picks out no preferred basis in the space of state vectors. The $2$-norm respects all unitary evolutions and does not restrict them. Think about it, if the admissible physical evolutions were restricted to discrete symmetries, e.g. permutations, there would be no continuity and no time as we know it. There would be, I daresay, no life as we know it and, the saddest of all, no beer to drink. A sobering thought.

I can easily fill few more pages with anecdotes about Peter for working with him was a truly unforgettable experience. Some of our numerous discussions, over a pint or two, resulted in joint papers, some in half baked ideas and some, like the one about the Born rule, were just pure fun. This is how we worked together. He has been an inspiring teacher in more than one way. Many thanks, Pete, and many happy returns.

\section{Jumping cats, phase space and atom-cavity dynamics -- Garraway}

Twenty-five years ago I started working as a post-doc with
Peter Knight at Imperial College.  At this time there was considerable
activity in Peter's group on the Jaynes-Cummings model.  In
particular, I remember Hector Moya-Cessa, Emerson Guerra, and Fernando
De Oliveira. (Later we were joined by Mike Rippin and Sergio Dutra,
with Bryan Dalton visiting at several points, as well.)  I was
impressed with the visualisation produced from the display of the
associated field Wigner functions. How far could this idea be pushed?
One of my earliest works with Peter addressed the topical issue of
quantum phase by using radial integration of the Wigner function to
obtain a probability distribution
\cite{PhysRevA.46.R5346,1402-4896-1993-T48-010}. Of course, we fully
expected any Wigner function negativity to present itself in some
interesting way, but it was surprising how well it worked. We explored
superpositions of Fock states, and the ubiquitous superpositions of
coherent states. (These were sometimes known as even and odd coherent
states, sometimes known as `cat' states, and sometimes known as
`kitten' states when the mean number of photons was low.) By using the
Wigner function to define a quantum phase distribution for these
various states, one also avoided the thorny issue of defining a phase
operator.

The visit of Gershon Kurizki brought ideas of creating interesting
quantum states through \emph{conditional measurements}
\cite{PhysRevA.49.535}. The process of conditioning states on
measurements was very logical, but only just becoming widely known at
that point \cite{Dalibard1992,Molmer1993,Carmichael1993}.  We used an
idealised Jaynes-Cummings model to explore this and we found that by
conditioning on measurements of the atom it was possible to place
coherent states around a circle in the cavity field phase space: this
could either make multiple component `cat' states, or in the limit of
overlapping coherent states a highly non-classical Fock state could be
made.
So we (including Gershon's team and Hector Moya-Cessa at Imperial)
were able to show how Fock states could be produced in an idealised
micromaser (i.e.\ atom-cavity system) simply through measurement and
by only accepting certain sequences of results. All this was nicely
visualised with help from the $Q$-function. Much later, it was amazing
for me to find that the Haroche's experiment in Paris used related
concepts, and went beyond these ideas (see
e.g.\ Ref.~\cite{sayrin_real-time_2011}).

The work on conditional sequences of measurements in the micromaser
coincided with an explosion of activity on `Quantum Jump' theory, on
various state diffusion schemes and on measurement interpretations of
all this. The reader may be aware that the quantum simulation methods
(such as quantum jumps) had emerged as numerical methods from the
challenges of simulating laser cooling \cite{Molmer1993}. However, their measurement
interpretations were very interesting as points of principle: i.e.\ the
numerical simulation could now be attached to a physical process. 
We used this approach to examine several atom-cavity problems taking
into account the effects of decoherence
\cite{PhysRevA.49.1266,PhysRevA.50.2548}. In particular, in
Ref.~\cite{PhysRevA.49.1266} we examined in detail the `jumping cats':
that is the time evolution of the even and coherent states when
conditioned on measurement results in the environment. We looked at a
conventional quantum jump scheme (linked to direct photo-detection) as
well quantum state diffusion (linked to a heterodyne measurement).
During the earlier work on phase distributions we had obtained a
numerical procedure for evaluating the Wigner function in the Fock
basis, which was very helpful in visualising the evolution of these
systems.
We looked at the jumps and evolution of Schr\"odinger cat states as
produced in a Kerr medium \cite{Garraway1996517}, but
the particular case of two-photon driving with a special two-photon
absorbing medium was of interest as it showed that it was possible for
decoherence processes to produce coherence such as that required for
the `cat' states. This was studied in several related works
\cite{PhysRevA.49.2785,PhysRevA.55.3842}, but the two-photon absorbing
medium required to generate equilibrium `cat' states this way was rather
exotic. However,
recently there has been work on superconducting qubits which uses a
related model \cite{10.3389/fict.2014.00001}: 
I would say this is an example of the recommendation that theorists
should not resist the temptation to explore what is thought to be
exotic at the time.

We looked into some more fundamental aspects of quantum jumps: for
example, if one just looked at the evolution of a state conditioned on
\emph{not} detecting a photon, then with energy supplied to the system
the time evolution of the state resulted in it being led to a
kind of `attractor' state which did not depend on the initial state
\cite{Steinbach1995}. This also played a role when we examined the
phenomenon of quantum jumps \cite{Garraway1995560}. A bit of
explanation is in order here as the phrase `quantum jumps' can be used
in two ways. We have been using it above in the context of conditional
measurement sequences, or in the unravelling of master equations,
where jumps in the wave-function take place at points of
measurement. However, `quantum jumps' also have a history in the
context of systems being `shelved' in a meta-stable state whilst
undergoing fluorescence on another strong transition
(see, for example,
\cite{PhysRevLett.57.1699,PhysRevLett.57.1696,PhysRevLett.54.1023}). 
What happens in this case is that when the atom, or ion, is `shelved'
the fluorescence stops abruptly---i.e. there is a quantum
jump. However, when we analysed this from the perspective of `quantum
jump theory' \cite{Garraway1995560}, we found that the system did not
jump into the metastable state: it time-evolved into it. Subsequently,
when leaving the metastable state it jumped out.
The Quantum Jump approach was sometimes implemented as a first-order
in time expansion of probabilistic processes: in
Ref.~\cite{PhysRevA.51.3302} we took this to a higher order in a
time-step $\Delta t$. This has the potential to make a more efficient
method for large problems.

The work on decoherence and jumps was reveiwed in several places
including Refs.~\cite{Knight1996,1402-4896-1998-T76-022} and
\cite{PlenioK98}.  
We used these techniques to analyse the situation of a three-level
atom in a cavity which resulted in atom-cavity quantum beats
\cite{PhysRevA.54.3592}. Later, this work developed into the full
pseudomode theory presented in
Refs.~\cite{PhysRevA.55.2290,PhysRevA.55.4636}
and many papers since then~%
\cite{PhysRevA.64.053813,PhysRevA.68.033809,PhysRevA.77.033831,PhysRevA.79.042302,PhysRevA.80.012104,LazarouJPB11,PhysRevA.86.012331}.

During my time with Peter, I also overlapped for a year with Nikolay
Vitanov who was working effects of pulses
\cite{PhysRevA.52.2245,0953-4075-28-10-010,VITANOV199531} 
and we managed to complete some work on time-dependent two level
systems \cite{PhysRevA.53.4288,PhysRevA.55.4418}, a topic I had become
interested in with my previous mentor Stig Stenholm.  This work
led to a tentative connection to Peter's activity in high harmonic
generation where, together with Florence Gauthey, we formulated an
extremely simple (two-level) model of the process
\cite{PhysRevA.56.3093}. Later, this work in the area of
time-dependent systems led me along a long path into another body of
work in the context of cold-atom physics and Bose-Einstein
condensation (recently reviewed in
Refs.~\cite{0953-4075-49-17-172001,Perrin2017181v2}). This is another
story but, in the end, it has come full circle as we currently work
on applications to rotation sensing in quantum technology.

I left Peter's group for a lectureship at Sussex University. The
work on cold-atom physics expanded. The
quantum optics work continued with my students and colleagues from
Imperial days such as Bryan Dalton. In quantum optics we work on
atom-cavity problems and the development of pseudomode theory.
It has been a pleasure to work with Peter and a privilege to spend
time at Imperial College---%
and, away from work, I have fond memories of Saturday lunchtime
gatherings at Chris and Peter's house which were always very
entertaining.

\section{Relativistic Quantum Optics and Extreme Fields -- Keitel}

During my time as a post doc in the group of Peter Knight from 1992 till 1997 he initially put me on the interesting project of entering the promising field of the interaction of matter with such strong laser fields that the dynamics becomes relativistic. We found that the relativistic dynamics gives rise to novel harmonic generation \cite{Keitel1} and to a breakdown of the stabilisation due to magnetic field effects \cite{Keitel2}. In what follows we put a major effort into the development of an efficient code to solve the Dirac equation in two dimensions \cite{Keitel3} powerful enough to evaluate the ionization in an extremely strong laser pulse. In addition I was involved in a number of other activities involving nonrelativistic  laser matter interaction and quantum optics which influenced clearly my future developments.

In the following steps I set up a group in Freiburg with a focus on relativistic laser ion interaction and then a division at the Max Planck Institute for nuclear physics in Heidelberg where my activities were broadened to include high precision QED, x-ray quantum optics and extreme field high energy laser physics.

If a very strong laser pulse impinges on a highly charged ion the interaction is influenced by relativity, QED
and nuclear effects. An example project considered how the tunneling process was affected by magnetic fields
and an adapted gauge invariant tunnelling picture was developed \cite{Keitel4}. In addition the non-adiabatic regime was
investigated \cite{Keitel5} and the tunnelling time was evaluated \cite{Keitel6} and just confirmed experimentally with our local colleagues in Heidelberg \cite{Keitel-e}.

Given progress in the development of high frequency laser sources we put forward methods how to treat the interaction of such laser fields with highly charged ions and nuclei. Here schemes for controlling nuclear dynamics \cite{Keitel7}, understanding astrophysical ion spectra \cite{Keitel8} and the tunneling time was evaluated \cite{Keitel9}. Finally my original interests in the control of the influence of spontaneous emission while with Peter \cite{Keitel10}, which was especially important for high-frequency transitions from atoms to nuclei, was continued with ideas how to reduce such unfavorable influences \cite{Keitel11,Keitel12}.

Finally considerable effort was put on understanding the interaction of matter with the strongest laser fields available and under construction. Here radiative reaction is a key issue \cite{Keitel13} and has become a major field of research along with the problem of pair production and other high energy processes \cite{Keitel14,Keitel15}. An interesting example activity was to show that laser-induced pair production can be reverted like in laser atom interaction such that a collider was put forward to show the feasibility of inducing high-energy collision processes plainly from illuminating vacuum with extremely intense but still feasible light sources \cite{Keitel16}. 

\section{Quantum walks for quantum information processing - Kendon}

When I moved to Imperial in 2000 to join Peter Knight's Theoretical Quantum Optics Group, the field of quantum information was gathering significant momentum, with seminal results coming from the group 
(e.g., \cite{letter} and \cite{murao2} -- \cite{JonathanPK00}).
With a computational physics background, my interest was soon caught by quantum versions of random walks \cite{aharonov00a,ambainis01a}, introduced to underpin quantum versions of classical randomised algorithms \cite{kendon06a}.  Typically, they provide a polynomial speed up, e.g., quadratic for quantum searching \cite{shenvi02a,childs03a}.  A quantum coin provides a richer range of behaviours than the classical ``heads'' or ``tails'' \cite{carneiro05a}, due in part to the generation of quantum entanglement.  We had lively discussions of the nature of the quantum contribution: in \cite{knight03b,knight03a} Peter highlights quantum mechanics as a wave mechanics, like classical optics.  A number of factors all need to be present to obtain a useful quantum advantage. Specifically, quantum walk algorithms are to be run on quantum computers, encoded into qubits, a subtlety I still regularly find myself explaining to physicists fifteen years later.  There is no meaningful speed up for a physical implementation of a quantum walk algorithm with a single walker, although such experiments are of fundamental interest, as Peter Knight and Barry Sanders explored \cite{sanders02a}, see also section \ref{Sanders}. However, if identical single photons are available, multiple photonic quantum walkers give rise to the BosonSampling problem \cite{aaronson10a}.  Calculating the permanent of a matrix in a linear optical setup was identified earlier by Peter and Stefan Scheel  \cite{scheel04a,PLK,aaronson10a}.  Approximating the permanent of a matrix is difficult classically but -- in principle -- easy using photons. However, efforts to  implement BosonSampling beyond two photons are experimentally very challenging: Peter Knight is involved in a collaboration currently underway \cite{ImperialOxfordProject} to achieve this.

Quantum walks continue to provide many diverse avenues for current research, I mention just two here that I have pursued since leaving Peter's group.  Inspired by the use of classical random walks to model the behaviour of the scalar matter field during the inflationary period of the early universe, discrete-time, coined quantum walks in one-dimensional potential wells provide an alternative model to consider.  A quadratic potential $V(x) = m^2x^2/2$ with a tuneable mass $m$, can be parameterised using a position-dependent biased coin operator $H$ with bias determined by $\theta$
\begin{displaymath}
H = \left(\begin{array}{cc} \sin\theta & \cos\theta \\ \cos\theta & -\sin\theta \end{array}\right)
\quad\text{where}\quad
\theta = \frac{1}{2}\arctan\left(\frac{m^2}{x}\right)
\end{displaymath}
is determined by the derivative (slope) of the potential at $x$. Choosing different functions for $V(x)$ allows for cubic, quartic, exponential and anything in between \cite{ohalloran17a}.

Quantum walks also feature as one pillar of \emph{continuous-time quantum computing} in which data encoded into qubits is evolved using an engineered Hamiltonian combined with coupling to a low-temperature environment.  The goal is to find a solution corresponding to the lowest energy eigenstate of the Hamiltonian.  The other pillar is \emph{adiabatic quantum computing} in which the same Hamiltonian components are applied in a slowly time-varying manner, keeping the quantum system in its ground state throughout.  Between these two extremes, a range of hybrid strategies allow the computation to be optimised for specific hardware characteristics, which will enable the maximum performance to be extracted from early quantum computers \cite{morley17a}. These ideas are contributing to the realisation of quantum computing as part of the quantum technologies programme that Peter has done so much to develop, and which funds my current fellowship.  Without his continuing support for my career, I would not be working in such an exciting and promising area of current research.

\section{Interaction of matter with strong laser fields -- Lein}
\subsection{Strong-field response of aligned molecules}
My research activity as a postdoc in Peter Knight's group at Imperial College 
in the years 2001/2002
focused on the response of small molecules such as H$_2^+$ and H$_2$ to intense-field irradiation.
Investigations of laser-aligned molecules in strong fields were relatively new at the time
and the first work on high-harmonic generation from adiabatically aligned molecules had been published
by the group of Jon Marangos in the year 2001 \cite{Velotta}.
Also, only few theoretical studies of the orientation dependence of high-harmonic generation
had been carried out \cite{Kopold,Lappas}.
We studied high-harmonic generation numerically for model systems and for arbitrary
angles between the molecule and the laser polarization \cite{LeinPRL2002,LeinPRA2002}.
We found that for a given orientation angle, the harmonic spectrum
exhibits a minimum at a position that is independent of the laser parameters.  
Similarly, the strength of a given
harmonic order depends on the orientation. A pronounced  minimum and 
a  sudden  jump  in  the  harmonic  phase  occur at a certain orientation angle.  
We performed a detailed analysis of this phenomenon and came to the conclusion that
two-center interferences within the molecule are responsible for the orientation-dependent signal.
Our discovery marked the beginning of research efforts in many groups to investigate
the response of aligned molecules theoretically and experimentally \cite{LeinJPhB2007}. 

Inspired by the possibility to use such effects for the imaging of molecular structure
and dynamics, we extended our studies to the orientation dependence of photoelectron momentum
distributions from strong-field ionization.
We computed the distributions of electrons from ionization of diatomic molecules in
linearly polarized pulses \cite{LeinPRA2002R}. We performed first the numerical solution of the time-dependent
Schr\"odinger equation and then set up a semiclassical model to describe the diffraction 
of electrons at the recollision with the parent ion. 
We found that diffraction is clearly visible in the electron angular distribution.
The possibility of recollision diffraction was anticipated already in 
earlier work on interference in molecular strong-field ionization
\cite{Zuo1996}. In the following
years, this method was confirmed experimentally and
became known as laser-induced electron diffraction \cite{Meckel2008,Blaga2012}. Today it finds increasing
popularity in molecular imaging strategies.

\subsection{Atoms and molecules in strong laser fields}
My group at the Leibniz Universit\"at Hannover works on a broad range
of strong-field phenomena in atoms and molecules. 
This includes the theory of photoelectron momentum distributions, high-harmonic generation
and time-dependent density functional theory.
Some of these activities can still be traced back to origins in the work done in the group of Peter Knight during
the years 2001/2002. For example, the imaging of molecular structure still plays a major role in current work,
see \cite{Petersen2015}.

One of the goals of high-harmonic generation is the generation of subfemtosecond pulses at high frequencies.
In this context, my group has recently collaborated with the Imperial College group to assess
the generation of attsecond pulses in the VUV range by appropriate filtering and by exploiting
elliptical polarization \cite{Henkel2013}.

Another current focus area of the work in my group is the response to two-colour fields, for example
$\omega$-$2\omega$ fields with orthogonal polarizations \cite{Eicke2017}. Possible applications are
the control of laser-induced dynamics by varying parameters such as relative phases and intensities
or the investigation of ionization dynamics by probing strong-field dynamics with an additional 
weak field.

The strong-field dynamics of matter remains an important area within the wider field of quantum optics.
The reliable numerical modeling of strong-field phenomena is a continuous challenge that must be 
taken on in order to keep up with the ongoing advances in experimental laser physics.

\section{Quasi-Probability for joint classical-quantum systems -- Milburn}
I joined Peter Knight's group as a postdoc in the summer of 1983 and
in 1984 gave a series of lectures, at the invitation of Peter Knight, on 
quasi-probablity distributions (QPD) in quantum optics. The concept of a QPD was 
first introduced by Wigner to give a phase-space description of quantum mechanics that mirrored classical
statistical mechanics. The Wigner distribution gives symmetrically ordered quantum moments by 
an integration over a phase-space probability distributions. New distributions were introduced in the 
1970s and 1980s  giving antinormally ordered moments (Q-function) and normally ordered moments (P-function) \cite{QIbook}. 
The Q-function exists as a positive distribution for all quantum states however the P-function does not exist as 
a regular function for many quantum states. Drummond and Gardiner introduced a generalised P-function that could 
respresent all quantum states using a regular positive phase-space distribution, albeit in double the ostensible dimension
of the underlying Hamiltonian system. 

In recent years, my group at The University of Queensland  has been considering 
how one might consistently combine quantum and classical descriptions. The motivation 
comes from the need to reconcile quantum mechanics with general relativity: perhaps one does not need to quantise gravity at all \cite{Jacobson, Verlinde}. 
Diosi \cite{Diosi} has proposed a consistent theory for the description of a hybrid quantum-classical dynamics. The key of Diosi's proposal is that a consistent treatment requires that noise be added to both the quantum and the classical system. In the case of the quantum system this necessarily introduces an intrinsic decoherence process. 

The fundamental problem of combining quantum and classical systems into a unified dynamical picture was pointed out long ago by Heisenberg. To see what goes wrong let us consider two free particles, one treated quantum mechanically and the other treated classically. Let $(Q,P)$ be the canonical coordinates for the classical system and let $(\hat{q},\hat{p})$ be the operators representing the canonical coordinates of the quantum system.  We will assume that the coupling is such as to provide a mutual constant force on each particle. To make matters simple we will assume that the coupling is 'impulsive' and only acts for a very short time during which we may neglect the free dynamics of the particles. We are thus treating the dynamics as a kind of `kicked system' in terms of a dynamical map. If we naively ignore all problems combining quantum and classical variables, the result of the coupling is the dynamical map relating the dynamical variables after the kick to the dynamical variables immediately before. 
\begin{eqnarray}
Q' & = & Q \label{dodgy-map1}\\
P' & = & P+g\hat{q}\\
\hat{q}' & = & \hat{q}\\
\hat{p}' & = & \hat{p}+gQ\label{dodgy-map2}
\end{eqnarray}
This map is generated by the interaction Hamiltonian $H_I=\kappa(t) Q\hat{q}$. The first, third and fourth equation can be given an immediate and unproblematic interpretation, but the second equation does not make any obvious sense as it maps a classical phase space variable into an operator on Hilbert space. However there is a more serious problem lurking in these equations. 

If this kind of coupling were possible we could use it to effect a violation of the uncertainty principle for the quantum system. By definition there is no problem in assuming the classical system can be prepared in a state so that before the kick occurs, the uncertainty in $Q$ and $P$ is zero. The last equation then tells us that no noise is added to the quantum systems momentum as a result of the interaction. Thus if we measure the momentum of the {\em quantum} system after the kick, with arbitrary accuracy, as there is no uncertainty in variable $Q$, we can infer (using the last equation) the momentum of the quantum system before the kick with arbitrary accuracy, i.e. $\delta p\rightarrow 0$.  On the other hand, we can also make an arbitrarily accurate measurement of the momentum of the {\em classical} system after the interaction and using the second equation infer the position of the quantum system before the interaction, again with arbitrary accuracy, $\delta q\rightarrow 0$.    We have thus an ability to infer both the position  and the momentum of the quantum particle before the interaction with arbitrary accuracy and a violation of the uncertainty principle is possible. This problem would not arise if both systems were treated quantum mechanically as then the quantum back action of one system on the other would prevent it. Clearly there is something quite wrong about the way we have combined a quantum and classical dynamical description. 

Having seen what might go wrong we can also see a way to fix the situation: we simply have to add sufficient noise to both the quantum and the classical system to ensure that the argument of the previous paragraph does not go through. If we add sufficient noise to the {\em momentum} of the quantum particle and also to the {\em momentum} of the classical particle before the map is completed the previous argument does not go through as we cannot invert the second and last equations to uniquely determine the position and momentum of the quantum particle before the interaction. Now, adding noise to a classical particle can be done quite straightforwardly without disturbing the position of this particle,  but adding noise to the momentum of the quantum particle comes at a price: it necessarily causes decoherence in the position basis. This is the core of Diosi's approach. 

There is no {\em fundamental} problem in making a quantum Hamiltonian depend on a classical coordinate.  For example in the map discussed in the previous paragraph we had $\hat{p}' = \hat{p}+gQ$ which is simply a classical displacement of the quantum momentum. 
The problems begin when we need to consider the possibility that the quantum system is reversibly coupled to the classical system. This is what has gone wrong in the second equation of the map, $P' = P+g\hat{q}$.  

In ordinary quantum mechanics when this becomes a problem we simply quantise the classical system. For example, in the case of an atom interacting with the electromagnetic field, there are indeed situations that require us to quantise the field. Fortunately there is a quantum theory of the electromagnetic field so there is no show stopper here. However in the case of a particle coupled to a gravitational field we are in trouble precisely because we do not yet have a viable quantum theory of gravity. In fact there is some doubt that a quantum theory of gravity is even an appropriate way to approach the problem in the first place \cite{Jacobson, Verlinde}. In the absence of recourse to a quantum theory of gravity, Diosi's self consistent phenomenological  approach would seem to be the only viable option. 

Can we give an explicit model for how Diosi's phenomenological approach works in the case of the simple map considered here?    One way to couple a quantum system to a classical system has been well known for many decades in the context of measurement based feedback. In the context of the present discussion this would amount to first measuring the position coordinate of the quantum system, i.e. measure $\hat{q}$ with result $\bar{q}$ and then condition a force on the classical system proportional to $\bar{q}$. This could be implemented by a classical Hamiltonian of the form $H_{fb}= \kappa \bar{q}Q$. We still need to include the effect of the classical system on the quantum,  $\hat{p}' = \hat{p}+gQ$,  so we will need to add in the term $H_{cl}= \kappa \hat{q}Q$.   

The model is based on generalising the unitary map previously described to a completely positive map. First let us define the state of the quantum+classical system before the map in terms of the factorised distribution over quantum states
\begin{equation}
\rho(Q,P)=\rho W(Q,P)
\end{equation}
where $\rho$ is the state of the quantum system given in terms of a density operator and $W(Q,P)$ is the state of the classical system given in terms of a phase-space probability density. The CP map is then defined by
\begin{equation}
\rho(Q,P)\rightarrow {\cal M}[\rho(Q,P)]
\end{equation}
where 
\begin{equation}
{\cal M}[\rho(Q,P)]=\int_{-\infty}^\infty d\bar{q} \hat{E}(\bar{q},Q)\rho\hat{E}^\dagger (\bar{q},Q)\exp\left [-g\bar{q}\frac{\partial}{\partial P}\right ]W(Q,P)
\end{equation}
with the Krauss operator $\hat{E}(\bar{q},Q)$ defined by 
\begin{equation}
\hat{E}(\bar{q},Q)=\left (2\pi\sigma\right )^{-1/4}\text{e}^{-(\hat{q}-\bar{q})^2/4\sigma+ig\hat{q}Q}
\end{equation}
which satisfies
\begin{equation}
\int_{-\infty}^\infty d\bar{q}\hat{E}^\dagger (\bar{q},Q)\hat{E}(\bar{q},Q)=1
\end{equation}

We define the reduced state of the quantum system after the map as 
\begin{equation}
\label{reduced-quantum}
\rho'=\int dQ dP \rho(Q,P)
\end{equation}
and the reduced state of the classical system after the map as 
\begin{equation}
W'(Q,P)={\rm tr}\left [\rho(Q,P)\right ]
\end{equation}
These can be explicitly calculated as
\begin{eqnarray*}
\rho' & = &  \int dQdP\  W(Q,P)\ \int_{-\infty}^\infty d\bar{q}\hat{E}(\bar{q},Q)\rho\hat{E}^\dagger (\bar{q},Q)\\
W'(Q,P) & = & {\cal E}[W(Q, P-g\bar{q})]
\end{eqnarray*}
where ${\cal E}$ means a classical average over the random variable $\bar{q}$.

This is the same as the  unconditional  state of the quantum system alone after a generalised measurement, with result $\bar{q}$,  defined by the corresponding Krauss operator. It corresponds to decoherence in the position basis and complementary noise added to the momentum and a random Hamiltonian displacement in momentum. The probability distribution for the measurement result $\bar{q}$ is 
\begin{eqnarray}
P(\bar{q}) & = & \int dQdP\  W(Q,P)\ \int_{-\infty}^\infty d\bar{q}\ {\rm tr}\left [\hat{E}(\bar{q},Q)\rho\hat{E}^\dagger (\bar{q},Q)\right ]\\
 & = & {\rm tr} \left [\hat{E}^\dagger (\bar{q},0)\hat{E}(\bar{q},0)\rho\right ]
 \end{eqnarray}

We can now check that this is an appropriate generalisation of the previous unitary map by computing moments. 
  \begin{eqnarray}
  \langle \hat{q}'\rangle & = & \langle \hat{q}\rangle\\
  \langle \hat{p}'\rangle  & = &  \langle  \hat{p}\rangle+g{\cal E}[Q] \\
  {\cal E}[Q'] & = &   {\cal E}[Q] \\
  {\cal E}[P'] & = &   {\cal E}[P]+g\langle \hat{q}\rangle
  \end{eqnarray}
The second order moments are given by
    \begin{eqnarray}
\langle \Delta \hat{q}^{'\ 2}\rangle & = & \langle \Delta \hat{q}^2\rangle\\
  \langle \Delta \hat{p}^{'\ 2}\rangle  & = &  \langle  \Delta \hat{p}^2\rangle+\frac{1}{4\sigma}+g^2\Delta Q^2 \\
\Delta Q^{'\ 2} & = &   \Delta Q^{2}\\
\Delta P^{'\ 2} & = & \Delta P^2 +g^2\langle \Delta \hat{q}^2\rangle + g^2\sigma
  \end{eqnarray}  
  These equations do not share the interpretational difficulties of those in Eqns. (\ref{dodgy-map1}-\ref{dodgy-map2}), but the model does imply quite new physics. Indeed Eq. (\ref{reduced-quantum}) implies decoherence in the position basis and complementary noise added to the momentum. It is in this manner that consistently combining quantum and classical mechanics necessarily leads to intrinsic decoherence 
  in the quantum sub system.

\section{Atom-field interaction in a cavity -- Moya-Cessa}
During the autumn of 1990 I joined  Peter Knight's group
as a PhD student with the idea of studying ways to generate
non-classical states of light by using light-matter interactions.
Soon we realized that the nearby off-resonant levels could affect
the realization of such non-classical states and allow us to
publish our first joint paper: `Power broadening and shifts of
micromaser lineshapes' \cite{KnightFirst}.

This paper treats the Jaynes-Cummings model (JCM) as a simplified
version of a more complex problem, namely, the interaction between
radiation and matter. The JCM models this interaction and uses the
so-called rotating wave approximation in order to produce a
solvable Hamiltonian. However, the approximation of the atom as a
two-level system may not explain some effects present in
micromaser experiments such as Stark shifts and asymmetries  in
lineshapes that may be attributed to the effects of nearby but
non-resonant levels. We have analyzed the effects of such shifts
and nearby levels on the collapse-revival dynamics of the
atom-field interaction \cite{KnightFirst}.

The effect of nearby and off-resonant levels was modelled in
\cite{KnightFirst} by incorporating to the Jaynes-Cummings
Hamiltonian a term describing the intensity-dependent shift of the
two-level transition caused by virtual transitions to off-resonant
levels. We were able to show that ac Stark shifts may strongly
affect the dynamics of the JCM.

Some years later we studied ion-laser interactions, where we were
able to show some interesting features related to the
quantized-field-two-level atom interaction. Namely, we showed that
the so-called Rabi model Hamiltonian may be written, by means of
a unitary transformation, as the ion-laser interaction Hamiltonian
\cite{KnightJonathan} (on resonance) and gave some families of
exact eigenstates for both problems. In particular it could be
shown that the (unnormalized) state
\begin{equation}
|\phi\rangle=\left(
\begin{array}{l}
\left(1+\frac{\omega_0}{2\omega}\right)|-{i}\frac{g}{\omega}\rangle+{i}\frac{g}{\omega_0}|{-{i}\frac{g}{\omega},1}\rangle\\
\left(1-\frac{\omega_0}{2\omega}\right)|-{i}\frac{g}{\omega}\rangle-{i}\frac{g}{\omega_0}|{-{i}\frac{g}{\omega},1}\rangle
\end{array}\right)
\end{equation}
is an eigenstate of the Rabi Hamiltonian $H=\omega
a^{\dagger}a+\frac{\omega_0}{2}\sigma_z-ig(a-a^{\dagger})\sigma_x$,
where the $\sigma$'s are the Pauli matrices and $a$ and
$a^{\dagger}$ the annihilation and creation operators for the
field, while $\omega$ is the field frequency, $\omega_0$ the
atomic transition frequency and $g$ the interaction constant. The
states $|{-{i}\frac{g}{\omega},1}\rangle$ are displaced number
states.

On the other hand, it has been known for a long time that there
exists an analogy between linear diatomic lattices and the Rabi
problem. Evanescently coupled waveguides have emerged recently as
a promising candidate for the realization of lattices with tunable
hopping that may emulate quantum mechanical systems such as the
atom-field interaction. In Ref. \cite{Actual} it has been shown
that they may also be used for the implementation of quantum and
classical discrete fractional Fourier transforms that may be of
interest for the quantum information community. These systems show
good perspectives, not only for quantum information purposes but
also for classical-quantum analogies as they share several
similarities to cavity-QED systems.

\section{Analysis of entanglement -- Murao}
I was a postdoctoral fellow in Peter Knight's group at Imperial from July 1996 to March 1999.  During that period, I worked on projects with
Peter in quantum optics and also in the then emerging field of quantum information.

In quantum optics, we worked on an analysis of decoherence in nonclassical motional states of a trapped $^9\mathrm{Be}^+$ ion \cite{murao2} in the breakthrough experiment by the NIST group reported in 1996 \cite{Meekhof1996}.  We compared the experimental results to predictions from a theoretical model that introduced sources of decoherence that destroy the characteristic coherent quantum dynamics of the system, but do not cause energy dissipation.  The sources of decoherence were introduced phenomenologically and then described using a macroscopic Hamiltonian representing an imperfect dipole transition and fluctuations of a vibrational potential.

I started my time in Peter's group focused on quantum optics, but  soon became fascinated by the work on quantum information that was starting to emerge in the group. I started to work in that area which has been the cornerstone of my research ever since. At that time, the operational properties of multipartite entanglement were not well explored comparing to bipartite entanglement.  Purification protocols of bipartite entangled states had been proposed, but they were not straightforwardly extendable to multipartite states.  In \cite{murao1}, we presented purification protocols for a wide range of mixed entangled states of many particles together with Vlatko Vedral and Martin Plenio in Peter's group and also with Sandu Popescu working at Cambridge at that time.   The protocols were one of the first multipartite entanglement purification protocols and have been useful for understanding multipartite entanglement.

I returned to Japan in April 1999 to take a postdoctoral position at RIKEN.  In October 2001, I was appointed as associate professor and then, in 2015, professor in the Department of Physics, the School of Science, the University of Tokyo.  I was the first faculty member to work on theoretical quantum information in the department, and also the first female associate professor or professor in the department.  In addition to my research, I have invested considerable time raising the profile in Japan of diversity, internationalization and the field of quantum information.

In my group at the University of Tokyo, I have been working on the analysis of entanglement and other non-local properties of quantum
mechanics and their applications for distributed quantum information processing \cite{murao3, murao4,murao5,murao6}.  Recently, I am also
interested in developing new quantum algorithms based on higher order quantum operations \cite{murao7}.

\section{Quantum computation with many-particle systems -- Pachos}
The year 2002, which I spent as a postdoc of Peter Knight at Imperial College London has been one of the most colourful and fast-paced years of my career. Peter encouraged me to work on quantum computation and optical lattices, a suggestion that has shaped my research up to now and started a wonderful journey that involved collaborations with theorists and experimentalists. Peter Knight's contributions to my scientific and academic life are immense and for that I am very grateful to him.

At the beginning of the new millennium, the focus of quantum computation and quantum information gradually moved from systems containing only a small numbers of particles to many-particle quantum systems. The theoretical understanding and the experimental implementation of controlling quantum systems with few qubits prepared the grounds for advances into many particle systems. Guided by Peter Knight I investigated possible applications of optical lattices in quantum information processing. Optical lattices, formed by Bose-Einstein condensate superposed by standing laser fields, can force thousands of atoms to arrange themselves in regular patterns. These patterns create ideal registers for quantum computation, where each atom encodes a single qubit. In 2003, we published the first proposal of quantum computation with optical lattices~\cite{Pachos2003}. 

In our paper~\cite{Pachos2003} we presented a dynamical control scheme to perform quantum computation on a one dimensional optical lattice. The model is based on atom tunnelling transitions between neighbouring sites of the lattice and the interactions between atoms within the same site. These operations can be activated by external laser beams resulting in a two-qubit phase gate or in an exchange interaction. The crucial element of our proposal is a two-qubit gate implementation based on controlled atom-interactions. Moreover, we proposed how this system offers itself to perform the three-qubit Toffoli gate with a single control operation, thus dramatically reducing the required resources. Using atom-interactions in optical lattices, as Peter suggested, allowed us to accurately manipulate many-particle systems for the storing and processing of quantum information~\cite{Scheel2006} and for implementing quantum simulations of condensed matter systems~\cite{Joo2007}.

Later on, I extended our initial work in different directions. First we tried to simplify the control procedures for manipulating quantum information in optical lattices even further by devising novel techniques based on the global addressing of encoded quantum information~\cite{Kay2004,Kay2006}. In this way, we bypassed technical difficulties which arise when each atom site needs to be addressed individually. We studied the efficient realisation of quantum computation of various geometries of optical lattices. In particular, triangular lattices helped to enhance global addressing techniques for performing quantum computation~\cite{Kay2005}. Moreover, honeycomb lattices could perform anyonic quantum computation~\cite{Pachos2006}. 

But the most exciting application of our work is in the area of quantum simulations. By now, optical lattices have proved to be an ideal platform for the simulation of exotic phases of matter. Using optical lattices, these can be realised in a controlled environment and their properties can be probed with sophisticated diagnostic tools. Initially, I studied triangular optical lattices for the generation of chiral phases~\cite{Cruz2005,Tsomokos2008} and three-body interactions~\cite{Plenio2004,Rico2004}. Our model was a direct but rather surprising result of our previous work on quantum computation on triangular optical lattices. Subsequently, I became more interested in the realisation of more exotic phases~\cite{Ripoll2007,PachosPhases2005,Guridi2002}. The sophistication of our newly-developed techniques even lead to the possibility of simulating relativistic quantum field theories \cite{Cirac2010} and Abelian and non-Abelian gauge theories~\cite{Maraner2009,Alba2013}. This surprisingly brought me back to my original research field of high-energy physics which can now be probed with tabletop experiments.

My most favourite application of optical lattices is the quantum simulation of topological systems. The possibility to realise exotic quasiparticles that exhibit anyonic statistics, which is unlike the statistics of bosons or fermions, is of fundamental interest but also  holds the promise for applications in quantum technologies. Our current investigations, which range from the study of topological phases of matter and their diagnostics~\cite{Lisle2014} to possible realisations with optical lattices~\cite{Palumbo2013,PachosMaj2013,Alba2011}, attracted numerous further theoretical investigations and have already been realised in the laboratory~\cite{Martini,Urlich}. Looking back, I feel ever so grateful for my interactions with Peter Knight that helped shape my academic and research life. 

\section{Atoms interacting with a broadband squeezed vacuum - Palma}
When I joined Peter Knight's group for three years as a visiting PhD student, the quantum optics community working on squeezed states began to be interested on the one hand in the correlations which characterise two modes squeezed states and on the other in the interaction between single atoms and a broadband squeezed vacuum. 
As mentioned in section \ref{BARNETT} one of the most peculiar properties of a two modes squeezed vacuum is the fact that each of the two modes has a thermal like reduced density operator while the correlations between the two modes are responsible of the reduced quantum noise in a field quadrature, at the expenses of an increased noise in the conjugate quadrature \cite{PLKRodney, thermoQO}.  A broadband squeezed vacuum is therefore a reservoir in a pure state showing, at the same time, phase sensitive noise and thermal like fluctuations. 
The joint presence of single mode thermal-like features and of inter-mode correlations, both due to the entanglement between the pairs of modes \cite{Adesso05}, are evident once we look at the expectation value of field operators. We recall that a broad band squeezed vacuum is defined as  
\begin{equation}
|0_S\rangle = \Pi_k \exp\{\eta^* \hat{b}_k \hat{b}_{-k}  -  
\eta\hat{b}^{\dagger}_k \hat{b}^{\dagger}_{-k}\} |0\rangle
\end{equation}
where $\hat{b}_k$  and $\hat{b}^{\dagger}_k $ are bosonic operators for sideband modes at frequency $ \tilde{\omega}_k = \Omega - \omega_k$ with respect to the squeezing carrier frequency $\Omega$ and $\eta = r\text{e}^{i\varphi}$ is a flat function of $\omega$, from which follows

\begin{eqnarray}
\langle \hat{b}^{\dagger}_k  \hat{b}_{k'}\rangle &=&\delta (\omega_k- \omega_{k'}) \sinh^2r \\
\langle \hat{b}_k  \hat{b}_{k'}\rangle &=&\delta (\omega_k+ \omega_{k'}) \cosh r \sinh r \text{e}^{i\varphi}
\end{eqnarray}

Crispin Gardiner \cite{Gardiner86} was the first to study the interaction between a single atom and a broad band squeezed vacuum. He showed that squeezed light, with its reduced electromagnetic field fluctuations in one quadrature phase,  can inhibit the phase decay of an atom. This gives three relaxation times: the usual longitudinal relaxation time and two different transverse relaxation times, which are inversely proportional to the variances of the two quadrature phases of the incident light. 
The same year Gerard Milburn analysed the energy-level shifts for a multilevel atom interacting with a squeezed vacuum \cite{Milburn86}. He found that the level shifts are made up of two contributions: the ordinary Lamb shift and a shift due to the squeezed-vacuum intensity spectrum (similar to the blackbody-radiation shifts). Our first contribution in this field was to unify the previous work using a simpler formalism valid for both resonant and non resonant interaction \cite{SingleAtom}. We did so using a set of Heisenberg equation of motion for the atomic operators in the limit of broadband squeezed quantum noise. We found that when the detuning between the atomic transition and the carrier frequency of the squeezing is large one recovers the results in \cite{Milburn86}, where the squeezing generates an atomic frequency shift dependent only on the occupation number. For zero detuning, neglecting the frequency shifts, we recovered the results of \cite{Gardiner86}. However we found that the decay of atomic coherences is more appropriately described by generalised quadrature dipole operators whose value depends also on the energy shifts and on the squeezing parameters.

The next question we addressed was under which conditions squeezing could affect non only the transverse but also the longitudinal decay constant. In other words we started wondering under which condition population decay could show phase sensitivity. It became clear to us that, as far as the population is concerned, the single two level atom decay is analogous to a decay in a heat bath at finite temperature due to the single photon nature of emission and absorption.  Two photon processes are however possible for two atoms cooperatively interacting with the field. We found that indeed this system shows phase sensitive population decay due to the cross correlations between the population and the polarisation dynamics of the two atoms \cite{TwoAtoms1}.  The most striking - and initially unexpected - result was the fact that the two atoms do not relax into a mixed state but rather in the following pure state \cite{TwoAtoms1,TwoAtoms2}

\begin{equation}
|\mathcal{S}(\vartheta )\rangle =  \exp [\vartheta/2( \hat{\sigma}_{-}^{(1)}\hat{\sigma}_{-}^{(2)} - \hat{\sigma}_{+}^{(1)}\hat{\sigma}_{+}^{(2)} )] |gg\rangle = \cos\left[\frac{\vartheta}{2}\right] |gg\rangle + \sin\left[\frac{\vartheta}{2}\right] |ee\rangle \end{equation}
where for the sake of simplicity we have chosen $\varphi =0$ and where
\begin{eqnarray}
\cos\left[\frac{\vartheta}{2}\right] &=&\frac{\cosh r}{[\cosh^2 r + \sinh^2 r ]^{1/2}}\\
\sin\left[\frac{\vartheta}{2}\right] &= &\frac{\sinh r}{[\cosh^2 r + \sinh^2 r ]^{1/2}}
\end{eqnarray}

Such pure state, known as two atoms squeezed states, is again characterised by entanglement between the two atoms, leading to a thermal like statistics for the reduced atomic 
density operators. This was one of the first examples of entanglement transfer due to a dissipative process and one of the first instances of pure state preparation through coupling with an engineered environment.

When two decades after the interest of the quantum information community was focused on geometric quantum computation, my collaborators and I came back to this system and we proposed a new way to generate an observable geometric phase by means of a completely incoherent phenomenon. The basic idea was to force the ground state of the system to evolve cyclically by adiabatically manipulating the environment with which it interacts. As a specific scheme, we indeed analyzed a multilevel atom interacting with a broadband squeezed vacuum bosonic bath whose squeezing parameters $r$ and $\varphi$ are smoothly changed in time along a closed loop \cite{Carollo1,Carollo2}. The interesting feature of such scheme is the the fact that the geometric phase imprinted on the ground state of the system is due to a purely incoherent engineered dynamics.

The late seventies and early eighties witnessed also a strong theoretical and experimental interest for the cooperative emission leading to the phenomenon of superradiance, characterised by a strong interplay between quantum and classical features for the atomic dynamics \cite{GrossHaroche}. On the one hand quantum fluctuations play a crucial role in the triggering of the superradiant decay by an ensemble of atoms initially all in their excited state, corresponding to the state of unstable equilibrium of an inverted collective Bloch vector. On the other hand, once the decay has started, the atomic dynamics can be described entirely in classical terms, and the quantum onset of the process manifests itself in large fluctuations of the macroscopic parameters characterising the decay in each single experiment. Superradiance can indeed be considered as a phenomenon of amplification of quantum fluctuations. 
Along this  line we studied the time evolution of an initially completely inverted assembly of atoms interacting cooperatively with a squeezed reservoir \cite{SuperSqueezed} and showed that the phase sensitive fluctuations of the electromagnetic squeezed vacuum field induce a distortion in the contour of the quasi-probability distributions of the atomic variables which is elliptical rather than isotropic in phase, as in usual superfluorescence. This distortion manifests itself through a phase dependence in the distribution of the delay time of the superradiant bursts. This  provides a macroscopic probe for a measure of the squeezing of the field.

Further features emerged when we studied the superradiant decay of two groups of atoms having slightly different transition frequencies and interacting with a squeezed electromagnetic reservoir \cite{SuperBeat}. Again the statistical properties of the onset of the decay process are modified with respect to those of usual superradiant beating because of the presence of squeezed field fluctuations. We showed that this influences both the auto-correlation properties of the whole intensity emitted by the two groups and the cross correlations of the emission processes of the two species of atoms. In particular we showed  that the intensity correlations between the two groups in the case of large detuning depend directly on the squeezing parameter.

Looking in retrospective it is clear that several of the themes which emerged in our research on the interaction between squeezed light and atomic systems, like environment engineering, bipartite entanglement, state purification, state engineering etc. have become of central importance in the quantum information arena which appeared on stage less than a decade after.

\section{Coherent laser-matter interaction -- Paspalakis}
I collaborated with Peter Knight initially as a PhD student under his supervision (October 1996 -- March 1999) and later as a postdoc in his group at Imperial College (April -- September 1999 and April -- October 2001). We even collaborated from a distance during my compulsory military service in Greece between my two postdoc periods. Peter Knight and I worked in the general area of quantum coherence and interference in quantum systems interacting with light. Our collaboration was roughly divided into five different topics: (a) coherent population transfer and quantum interference in quantum systems with electronic continua, (b) novel optical effects in quantum systems with vacuum induced coherence, (c) electromagnetically induced transparency and coherent pulse propagation beyond three-level systems, (d) laser-induced atomic localization, and (e) quantum optics in atomically doped photonic crystals. Below, there is a short summary of some of the effects that we studied together and some connection with my current research activities. I have the greatest memories from our collaboration. Peter is a prolific physicist, a great mentor and an excellent collaborator, who gives you the opportunity to work in really interesting problems and allows you to evolve scientifically. He is also a kind, warm, polite and truly remarkable person. The interaction with Peter certainly determined my scientific career and also shaped me as a person as well. I am really happy to contribute this part in honour of his 70th birthday as a small token of appreciation and wish him many happy and fruitful years to come. 

The study of the effects of dissipative processes in coherent laser-matter interactions is a
well-established area of research in quantum optics. In the past it was generally thought
that dissipative processes destroy quantum coherence. However, there are situations in
which dissipative processes induce rather than destroy quantum coherence. A prerequisite
for this is the presence of interfering decay mechanisms. A prototype system is the
ionization of an atom near an autoionizing resonance, the well-known Fano system \cite{Fano61a}. In this case, as Fano analyzed
in his seminal work, ionization occurs via two indistinguishable paths; a direct one and an indirect resonant one via the autoionizing state. These paths
interfere and the interference leads to several interesting phenomena, such as for example,
the modification of the ionization profile induced by a weak laser field and coherent
population trapping \cite{Knightrev}. A similar system where such interferences occur is the laser-induced continuum structure (LICS) scheme \cite{Knightrev}.
In this case the interference is induced by two
laser fields each coupling a bound state with the same continuum of states. Interference
can be also induced through the process of spontaneous emission \cite{Keitel12}. An example
illustrating this phenomenon occurs when two close lying excited states interact with
the same vacuum field \cite{Paspalakis00a}.

In the phenomenon of LICS a strong laser pulse embeds a discrete state in the continuum \cite{Knightrev}. If a weak laser pulse
is used to probe this embedded state, a Fano-type profile \cite{Fano61a} is observed in the
photoionization rate. If the probe laser is strong then population trapping can be achieved
under specific conditions \cite{Knightrev,PaspalakisJPB98a}. In a clear experiment of LICS in helium, Halfmann {\it et al.} \cite{Halfmann98a} used a Nd:YAG laser
to embed the 1s4s $^{1}$S$^e$ Rydberg state in the continuum and the resulting modification of the
1s2s $^{1}$S$^e$ photoionization yield was measured. The fact that the Fano asymmetry parameter $q$
for this system is small facilitated the observation of a deep minimum in the photoionization
spectrum of the 1s2s $^{1}$S$^e$ state. Together with Niels Kylstra, Peter Knight and I gave an independent theoretical description
of this LICS process in helium, using the non-perturbative R-matrix Floquet approach \cite{PaspalakisLICS1,PaspalakisLICS2} and found excellent agreement with the experiment. We also investigated the regimes where the incoherent channels due to multiphoton processes become significant
and examined the LICS due to the 1s4d $^{1}$D$^e$ state which could be observed within the range
of the experimentally achievable probe laser frequencies. The validity of
the simple two-level models that are frequently used for modeling LICS was tested by comparing
these results with R-matrix Floquet calculations.

The effects of smooth laser pulses were investigated by us with Markos Protopapas in both an autoionization system
and in the LICS scheme \cite{PaspalakisJPB98a}. We were particularly interested in the
phenomenon of coherent population trapping in these systems and discussed the effects of time-dependent
pulses on this trapping. We presented analytical solutions to both systems which hold for any
pulse shape, provided the adiabatic approximation holds. We also paid special attention to the
effects of large pulse energies on the trapped population, which were found to be detrimental and
proposed a possible solution to this problem by using chirped (time-dependent) frequencies. For a counterintuitive sequence of properly chirped laser pulses, the LICS scheme was also shown to lead to very efficient population transfer between the two bound states via the continuum of states \cite{Paspalakis97a}. Moreover, Peter Knight and I with Niels Kylstra studied the propagation dynamics of a short laser pulse interacting with an autoionizing medium \cite{PaspalakisFano99}.
We showed that Fano interference can lead to transparency in the medium, thereby allowing the laser pulse to
propagate without absorption. We placed emphasis on establishing the connection between adiabatic
population trapping in short, pulsed laser fields \cite{PaspalakisJPB98a} and transparency in the propagation of the laser pulse in the
medium \cite{PaspalakisFano99}. Furthermore, we calculated both analytically and numerically corrections to the adiabatic behavior and showed that in the first approximation the laser pulse retains its shape and only its group velocity is modified.

Peter and I also presented several studies on novel optical effects arising from quantum interference in spontaneous emission \cite{Paspalakis98a,Paspalakis98b,Paspalakis99a,Paspalakis00b}. Specifically, we studied a four-level quantum system with two closely lying upper states that leads to quantum interference in spontaneous emission under excitation with two laser fields with equal frequencies and different phases between them that couple the ground state to the two excited states and showed the potential for coherent control in the optical response of the system \cite{Paspalakis98a}. Besides partial or even complete coherent population trapping, which is possible in this system under proper excitation, we also showed extreme spectral narrowing and selective and total
cancelation of fluorescence decay as the relative phase of the two laser fields was varied. Christoph Keitel, Peter Knight and I discussed the spontaneous emission from a coherently prepared and microwave-driven doublet of potentially closely spaced excited states to a common ground level \cite{Paspalakis98b}. Multiple interference mechanisms were identified
that may lead to fluorescence inhibition in well-separated regions of the spectrum or act jointly in cancelling the
spontaneous emission. In addition to phase-independent quantum interferences due to combined absorptions
and emissions of driving field photons, we distinguished two competing phase-dependent interference mechanisms
as means of controlling the fluorescence. The indistinguishable quantum paths may involve the spontaneous
emission from the same state of the doublet, originating from the two different components of the initial
coherent superposition. Alternatively the paths involve a different spontaneous photon from each of two
decaying states, necessarily with the same polarization. This makes these photons indistinguishable in principle
within the uncertainty of the two decay rates. The phase dependence arises for both mechanisms because the
interfering paths differ by an unequal number of stimulated absorptions and emissions of the microwave field
photons.

Furthermore, we showed that interference arising
from spontaneous emission can make either a four-level medium \cite{Paspalakis99a} or a three-level V-type medium \cite{Paspalakis00b}
transparent to a short laser pulse. We have shown that under
specific conditions, namely, the conditions that lead to a dark
state in the respective system, the media can become transparent to
the laser pulse when the evolution of the system is adiabatic.
First order non-adiabatic effects change only the group velocity of the pulse without leading to absorption. Stronger non-adiabatic effects or extra decoherence effects may lead to absorption and changes in the group velocity and pulse shape.
We also showed that the probe absorption spectrum of a weak field coupling one transition of a $\Lambda$-type system shows transparency when the other transition
is coupled to a strongly modified photonic reservoir, like when decaying spontaneously near the edge of a photonic band gap material which is modeled by an isotropic dispersion relation \cite{Paspalakis99b}. This transparency occurs even in the presence of the background decay of the upper quantum level. For the types of transparency discussed above no additional laser field is required. This fact distinguishes these works from propagation studies that are based on electromagnetically induced transparency (EIT) \cite{EITrev}.

Both quantum interference effects in spontaneous emission as well as strong coupling effects via spontaneous decay may occur in quantum emitters near plasmonic nanostructures, a field that has attracted significant attention recently \cite{Qplasmonsrev}, and my group at the University of Patras is currently involved. For quantum interference effects to occur in quantum emitters in free space
the corresponding electric dipoles for the spontaneous emission process should
be nonorthogonal, with the maximum quantum interference effects occurring for parallel (antiparallel) dipoles. This is a condition that is
not easy to meet in real systems. However, the spontaneous emission of a V-type quantum emitter with orthogonal electric dipoles for the transitions of the upper states to the lower state may exhibit interference effects near a plasmonic nanostructure \cite{Yannopapas09a,Thanopulos17a}, since the plasmonic nanostructure affects differently the electromagnetic
modes along mutually perpendicular directions, a phenomenon known as the anisotropic Purcell effect \cite{Keitel12,Agarwal00a}. This way the effects of quantum interference in spontaneous emission are efficiently simulated. A quantum system that has shown remarkable optical response is a four-level double-V-type quantum system that exhibits quantum interference in
spontaneous emission when placed near a two-dimensional array of metal-coated dielectric nanospheres. In this quantum system, one V-type transition
is influenced by the interaction with localized surface plasmons while the other V-type transition interacts with free-space vacuum and with the
external laser fields. When this system interacts with a weak probe laser field, it leads to optical transparency accompanied with slow light
\cite{Evangelou12a} and strongly modified Kerr nonlinearity \cite{Evangelou14a}. Additionally, when the system interacts with two fields it
leads to complete optical transparency, gain without inversion, huge enhancement of the
absorption at the central line, and a phase-dependent absorption spectrum \cite{Paspalakis13a,Carreno17a}. Strongly modified resonance fluorescence can also occur for the same four-level double-V-type quantum system driven by two laser fields when placed near a two-dimensional array of metal-coated dielectric nanospheres, where both the resonance fluorescence
spectrum and the second-order correlation function are strongly phase-dependent so that the relative phase of
laser fields can be used for the efficient control of the resonance fluorescence characteristics \cite{Carreno17b}. We note also that recently,
phase-dependent optical absorption and phase-dependent fluorescence of molecules near metallic nanostructures
have been experimentally observed \cite{Pirruccio16a}.

Peter and I also studied quantum coherence and interference effects in quantum systems that do not involve quantum interference of decay channels. A very interesting and potentially useful quantum optical phenomenon is that of EIT \cite{EITrev}. The prototype system for EIT is a three-level $\Lambda$-type quantum system which interacts with a weak probe field and a coupling field. The application of the coupling field renders the system transparent to the probe field which is also accompanied with slow light. We extended the EIT phenomenon beyond three-level systems and studied EIT in a four-level quantum system in a tripod
configuration \cite{Paspalakis02a}. We have shown that the system can exhibit double transparency windows and,
in general, the group velocity of the probe laser pulse can obtain at most two different values at transparency. These group velocities can be controlled by varying the Rabi frequencies of
the coupling laser fields. We also extended EIT in a more general $(N+1)$-level quantum system with $N$ lower levels and one upper level which interacts with
$N$ laser fields \cite{Paspalakis02b}. Taking that the system is initially prepared in a particular lower level we studied the absorption
and dispersion properties of a probe laser field coupling this level to the upper level. We have shown that the
system can exhibit multiple transparency windows. At most $N-1$ transparency windows can occur and, in general, the
group velocity of the probe-laser pulse can obtain at most $N-1$ different values at transparency. These group velocities
can be controlled by varying the Rabi frequencies of the coupling laser fields. These effects have been experimentally observed \cite{Wilson05a,Wang06a,Gavra07a,Han08a,Yang15a} and can be important in quantum information processing \cite{Rebic04a}.

Finally, Peter and I proposed a method for subwavelength localization of an atom using quantum interference \cite{Paspalakis01a,Paspalakis05a}. We used a three-level
$\Lambda$-type atom that interacts with two fields, a probe laser field
and a classical standing-wave coupling field. If the probe
field is weak then the measurement of the population in the
upper level can lead to subwavelength localization of the
atom during its motion in the standing wave. The degree of
localization is dependent on the parameters of interaction,
especially on the detunings and the Rabi frequencies of the
atom-field interactions. These effects has been experimentally verified \cite{Proite11a,Miles13a,Miles15a}.

\section{From The JCM To Entanglement - A Midsummer Knight's Dream? -- Phoenix}
It's never a good idea to stretch analogy, metaphor or simile too far, but
when I joined Peter Knight's group as a somewhat gauche doctoral student at
the tail end of what passes for Summer in the UK back in 1986 I wondered
whether I had stepped right into the pages of Shakespeare's play. Peter, like
Oberon, dazzled us all with his power, his great passion, his boundless energy
and extraordinary insight. I was in awe and, truth be told, I still am.

I had completed my undergraduate degree in Theoretical Physics at York in the
UK and was fortunate to have turned my third year project into a modest
publication. I'd used entropy to characterize the behaviour of a stochastic
nonlinear system in which the probability distribution showed behaviour akin
to phase transitions. Peter got me started on looking at the celebrated
Jaynes-Cummings model. It's a great place to learn some of the basics of
quantum optics within a simple, and exactly solvable, model. I don't think he
expected us to find anything new, at least not immediately, but he wanted me
to continue my work with entropy with a view to exploring any connections
between the complicated and random-looking evolution of the JCM with coherent
state fields, and notions of quantum chaos. 

But we did find something new and unexpected, and little did we know how
important notions of entropy in quantum optics and quantum information would
become. The entropy showed us, amongst other things, that the celebrated
revival phenomenon of the JCM was heralded by a partial disentanglement of the
atom and field \cite{Annals}. Furthermore, by explicitly solving for the field
states, we found that the field dynamics could be entirely charaterized by
just 2 quantum states. It was initially puzzling until we found the work of
Araki and Lieb on quantum entropies \cite{Araki}. They had shown that for any
two quantum systems $A$ and $B$ their entropies are bounded by the remarkable
inequality%
\[
\left\vert S_{A}-S_{B}\right\vert \leq S\leq S_{A}+S_{B}%
\]
The RHS is of course just the familiar classical bound; the total entropy of 2
systems must be less than or equal to the sum of the individual entropies. It
is the LHS of this inequality that is profound and remarkable and purely
quantum in origin. For pure states of 2 quantum systems the entropies of the
individual systems are always equal! 

By happy coincidence Steve Barnett was working as a postdoc in Peter's group
and he'd been looking at thermofields. On a long walk on a rainy night in
Hanover after a large conference dinner, we thrashed out the details of what
we were to term the `index of correlation'. Previously, I'd been using entropy
mostly as a statistical measure rather than explicitly drawing on its
connection with information. Memory is a fickle mistress and my recollection
differs slightly from Steve's in that I felt we'd hit upon something that
could be of some importance and value. I felt that our information-theoretic
approach \cite{Simon1} which not only characterized correlation but also
allowed to us to demonstrate that 2-mode squeezed states were the most
strongly correlated possible for two field modes, subject to a constraint on
the mean energy, was perhaps aiming towards something profound.

Our subsequent work showed that quantum systems can contain twice as much
information in their correlation as their closest classical counterparts
\cite{Simon2} and that measurements on the individual systems alone could only
recover half of the available information at most; to recover all of the
information joint measurements are necessary \cite{Phoenix1}.

Three decades later we can see that notions of entanglement and entropy have
become central to the field of quantum information, but I still occasionally
wonder whether we haven't missed a trick or two. The `index of correlation',
or more properly the mutual information, is one of many ways to characterize
entanglement, yet as a measure of \textit{correlation}, it is fundamental. A
measure of the overall correlation between quantum systems should satisfy 3
properties. It must (a) be positive (b) be basis-independent \ and (c) it must
be additive for independent (non-correlated) quantum systems so that

\begin{description}
\item[(i)] $I\left(  \rho\right)  \geq0$

\item[(ii)] $I\left(  \rho\otimes\sigma\right)  =I\left(  \rho\right)
+I\left(  \sigma\right)  $
\end{description}

These conditions are sufficient to specify the familiar quantity $I\left(
\rho_{A},\rho_{B}\right)  =S\left(  \rho_{A}\right)  +S\left(  \rho
_{B}\right)  -S\left(  \rho_{AB}\right)  $ as the unique measure satisfying
these properties. As \textit{purely a measure of overall correlation}, then,
the mutual information is, I think, rather fundamental. Its relationship to
entanglement and `non-classicality' is more complicated. With this measure of
correlation it is clear that the maximally correlated state of $n$ qubits has
a correlation strength of just 1 bit greater than the maximally correlated
state of $n$ classical bits; for qubits the difference between `classical' and
`quantum', in terms of correlation strength, amounts to just 1 bit. This is
related to the ability to purify the mixed state $\left\vert 000\ldots
0\right\rangle \left\langle 000\ldots0\right\vert +\left\vert 111\ldots
1\right\rangle \left\langle 111\ldots1\right\vert $ with the addition of just
1 extra qubit \cite{Phoenix}.

It all seems quite removed from those heady days spent poring over the JCM,
yet in that deceptively simple model the origins of some subsequently
important ideas can be found, at least partially. Peter tended to eschew the
overly technical approaches, not from any lack of ability, but I think more
because of the towering physical intuition which is at the heart of his work.
I don't know whether it was this intuition that saw something of value in
quantum entropies applied to quantum optics, I suspect so, but I remain
grateful for it and, yes, still in awe of it. 

The time I spent in Peter's group was one of the most exciting times of my
life and, quite frankly, I still miss those days. Peter is one of the most
inspiring and remarkable men I have ever met. Peter, it was an honour and a
privilege; thank you and Happy Birthday. Long may Oberon reign!

\section{Multiphotonic excitation -- Piraux}
\maketitle
I did my postdoctoral research with Peter Knight from 1986 to 1990. This research has been devoted to the theoretical study of the interaction of atoms with intense laser pulses with a strong focus on various  threshold effects. Having a quantum collision theory background, this field was new for me. I would like to warmly thank Peter for trusting me and for giving me the opportunity to take advantage of his expertise in this field. \\

We started with a detailed study of the near-threshold excitation of continuum resonances \cite{piraux90}. Such process combines the irreversible decay characteristic of all continuum couplings with the establishment of coherent superpositions, Rabi oscillations and population trapping more usually associated with bound-state dynamics. These latter effects are expected when the coupling strength from the initial discrete state to the continuum becomes as large as the width of the continuum density of state distribution, or approaches the energy distance from the threshold to the part of the continuum excited. In order to treat such process we used and compared several theoretical  alternative descriptions including a non perturbative coupling of a discrete state to a structured continuum with a threshold, a complete Fano diagonalization of the interacting system and a dressed-state picture.

 In this model-based study, it had been assumed that, for photon energies above the ionization potential,  the presence of various Rydberg  series that converge towards the ionization threshold is irrelevant. To analyze the specific role of these series when the photon energy is just below the ioni\-zation potential, we have, in the first step, examined the near-threshold photoionization of atomic hydrogen, in its ground state, and exposed to a monochromatic electric field \cite{piraux89}. We have demonstrated that these Rydberg series generate a very rich structure in the photoelectron energy spectrum through the action of the counter-rotating wave dynamics. This very rich structure does not involve any continuum-continuum transition and is important for photon energies lower than the ionization potential but much higher than the ponderomotive energy. In the second step, we have assumed that the electric field is sharply turned on and that the hydrogen atom is initially is the 2s state. We have shown that in addition to the rich structure that the counter-rotating dynamics generate in the electron energy spectrum, a resonance structure close to threshold appears, which results from non adiabatic transitions between degenerate opposite-parity states of atomic hydrogen \cite{millack90}.
 
For very strong peak intensities and photon energies lower than the ioni\-zation potential, the presence of the Rydberg series leads to a fascinating and counter-intuitive phenomenon namely the suppression of atomic  ionization. At very high frequencies, far above the ionization potential, this ionization suppression had been linked with the emergence of a `dichotomous'' wave function reflecting the nearly classical oscillation of the atomic electron in an ultra-intense field, the strength of which is of the order or higher than the atomic unit. In these conditions, the electron stays most of its time far away from the nucleus, in a region where the effective interaction with the external electric field is very weak since a quasi-free electron cannot absorb or emit photons. For frequencies lower than the ionization potential, we have shown that such ionization suppression can actually be caused by the creation of a spatially extended wave packet through Raman transitions from the ground state via high-lying Rydberg and continuum states to a superposition of Rydberg states with small initial overlap with the nucleus \cite{burnett91}. This suppression mechanism differs from other mechanisms involving population trapping where suppression results from the destructive interference of transition dipoles.

Since my postdoc at Imperial with Peter Knight, research on laser-matter interactions has evolved in two main directions. On the one hand, the develop\-ment of the free electron lasers has allowed one to study the interaction of complex systems (atoms, molecules and clusters) with XUV radiation. On the other hand, rapid progress in high order harmonic generation has led to the production of ultrashort pulses whose duration, a few tens of attoseconds, is of the order of the characteristic time of electron dynamics in atoms and molecules. The interaction of such ultrashort pulses with an atom opens the door  to real-time observation and time-domain control of atomic-scale electron dynamics with an unprecedented sub-atomic time and spatial resolution. The research I have performed and I am still performing at the Universit\'e catholique de Louvain (Belgium) follows these  two new directions. However, I would like to finish this short contribution by briefly discussing an issue, related somehow to what I have done at Imperial and which is back on the front stage, namely the mechanism of excitation of an atom in its ground state to a coherent superposition of  Rydberg states in the quasi-static field limit (frequency going to zero while the number of optical cycles within the pulse is kept constant). In 2008, Nubbemeyer {\it et al.} \cite{nubbemeyer08} observed, deep in the tunneling regime, a substantial fraction of helium atoms in excited states after having been exposed to a strong 800 nm laser pulse. As their findings are compatible with the strong field tunneling-plus-rescattering model, they concluded that the excited state population trapping is predominantly due to a recombination process that they called frustrated tunneling. On the other hand, Beaulieu {\it et al.} \cite{beaulieu16} studied recently the spectral, spatial and temporal characteristics of the radiation produced near the ionization threshold of argon by few-cycle 800 nm laser pulses. They showed that atomic resonances lead to the strong enhancement of harmonics just above the ionization threshold, thereby emphasizing the multiphotonic character of the excitation process.

In order to elucidate the actual excitation mechanism, my group in Louvain has proceeded in two steps. We have first studied numerically the excitation probability for a broad range
of values of laser parameters. In the second step, starting from first principles, we have derived analytically the quasi-static field limit  of the excitation probability. We have found that the excitation mechanism is predominantly of multiphotonic character and showed that the quasi-static field limit of excitation probability depends strongly on the carrier phase even when the number of optical cycles within the pulse is large.

\section{Quantum Information Science: From Entanglement Theory to Applications in Biology \& Medicine -- Plenio}
{\em Introduction ---} In this contribution I briefly explain how I had come to join the 
group of Peter Knight Knight as a postdoc in 1995 and combine it with an outline of
the topics that I have been working on in my postdoc as well as an outline of my career since
then.

{\em From G{\"o}ttingen to London ---}
In 1994, Peter held a Humboldt Research Award at Konstanz University and used it to
visit German research groups. He came to give a seminar in the Department of 
Theoretical Physics at G{\"o}ttingen University where I was doing my PhD. 
After walking him across town and letting him educate me about the founder of G{\"o}ttingen
University, King George II, a Hannoverian, I took the opportunity to ask him for 
possibilities to join his group at Imperial College - rather fitting as it 
had been founded by Queen Victoria, the great granddaughter of George II, and 
her husband the German Prince Albert. Despite my lack of history knowledge he 
offered me an appointment.

{\em Imperial College London: Postdoc 1995 - 1997 and Staff 1998 - 2009 ---} Supported by a 
Feodor Lynen fellowship of the Humboldt Foundation I arrived in Peters's group on January 
10, 1995 and began to work in a variety of directions. One avenue that I became interested 
in was the theory of entanglement, an area of investigation that is today embedded in the 
larger field of resource theories. In a nutshell, a resource theory considers constraints 
that are imposed on us in a specific physical situation. In entanglement theory this is the 
practical inability to perform joint quantum operations between distant laboratories. This 
restricts us to local operations and classical communication. Executing general quantum 
operations under such a constraint then requires the consumption of quantum states that
contain a relevant resource, e.g. entangled states, that are provided to us at a certain cost. 
Within such a framework one then wishes to establish which states contain the resource 
(characterisation), to quantify how much of the resource is contained in such a state 
(quantification), how one can interconvert resource states into each other under the given 
contraints (manipulation) and finally how to measure the resource efficiently in experiment 
(verification). Our first work in this area quantified entanglement by distance based entanglement 
measures, most notably the relative entropy of entanglement and formulated and promoted an
axiomatic approach to entanglement quantification \cite{VedralPR+97,VedralP98}. Over the years, 
then as Imperial College member of staff, the exploration of entanglement theory lead amongst 
others to the discovery of the concept of
entanglement catalysis \cite{JonathanP99}, the logarithmic negativity as an entanglement measure 
\cite{Plenio05} as well as the relationship of entanglement theory to thermodynamics \cite{BrandaoP08}. 

But I had always been interested in implementations which has formed another important facet of my 
work throughout my research career, initially focussing mainly on quantum optics. In this 
area Peter and I wrote a review article on the quantum jump approach \cite{PlenioK98}
that is still being in use today. This method for treating the system-environment interaction 
also played a role in the analysis of the limitations of quantum information processors due 
to their interaction with the environment. Focussing this analysis on trapped ion implementations,
we derived bounds on achievable gate fidelities. We pointed out that in laser driven gates 
the off-resonant coupling to energy levels
outside of the qubit manifold would eventually become relevant and impose gate errors
in the $10^{-4} - 10^{-6}$ range \cite{PlenioK96,PlenioK97}, then considered a small effect 
but today starting to be of relevance in actual implementations \cite{BallanceHL+16}.
The theme of system environment interactions continues to play a significant role in my
work and while at Imperial it led to the first proposals, jointly with my collaborator Susana 
Huelga, for using noise and dissipation to generate entanglement \cite{Martin}
and eventually led to the idea of examining the interplay of noise and coherence in biological 
systems \cite{PlenioH08} which is work that we continue to pursue in our new home at Ulm University.

After my postdoc, I continued my academic career at Imperial College as a member of staff. 
Indeed, at the beginning of 1997 I had made the resolution that I would leave physics to 
seek the riches of the banking world should I not be able to secure a permanent academic 
appointment, i.e. a Lectureship, somewhere in the UK by the end of that year. Just in time, 
in the autumn of 1997, a Lectureship suitable for a theoretician came up in the Optics Section 
and it was fortuitous that Peter was able to convince the selection committee that I was the 
man for the job. Thus, instead of a life of wheeling and dealing in the City of London I was 
set up for a career in science, initially at Imperial College (1998 - 2009) and since 2009 as
a Humboldt Professor in Ulm.

{\em Ulm University 2009 - 2017 ---} At Ulm, with the group that I have set up with Susana Huelga, 
we continue to pursue some of the topics described above and added others to the mix.

In the area of resource theories we are now pursuing various avenues but most notably
perhaps we have introduced the resource theory of coherence \cite{BaumgratzCP14} whose 
connections to thermodynamics and to entanglement theory are now a very active field of
research. In the study of system-environment interactions we have taken a rather different
turn and are now pursuing the study of quantum effects in biological systems \cite{HuelgaP13}. 
Here we identified and highlighted the central role of the coupling between electronic motion 
with vibrational degrees of freedom in this field \cite{PriorCH+10,ChinDC+10,ChinPR+13}. Our 
contributions to the experimental verification of these effects even in organic photovoltaics 
\cite{deSioTR+16} demonstrate that this concept has a role to play in the understanding of 
charge separation dynamics and other physical processes.

We remain interested in the effect of noise on quantum information processors but now we are 
looking at coloured noise whose effect can be suppressed by dynamical decoupling methods. This 
is relevant e.g. for microwave-driven trapped ion quantum information processors that are based 
on the use of magnetically sensitive states which are affected by magnetic field fluctuations. 
In work from my time at Imperial College we had realised that continuously driving a qubit 
transition it was possible to create a fast quantum gate which, while protected against coloured
dephasing noise, was still rather sensitive to power fluctuations in the driving fields 
\cite{JonathanPK00}. Building on this we showed that this problem can be circumvented by using 
more energy levels and thus creating a very well protected microwave driven quantum gates 
\cite{TimoneyBJ+11}. We are applying similar methods to colour centers in diamond, whose 
electron spin is subject to magnetic noise from both impurities and nuclear spins in the bulk and
the surfaces which have enabled us to contribute to the development of sensors that are
able to detect individual nuclear spins outside of diamond \cite{MullerKC+14} and even 
perform nanoscale NMR \cite{SchmittGS+17}. Thinking about these kind of problems has also 
led us to realise that colour centers in diamond provide a potential tool to improve MRI (Magnetic Resonance Imaging) by using them as a resource of nuclear hyperpolarisation, that is polarisation that 
far exceeds the thermal equilibrium value, of nanodiamonds or liquids above the diamond 
\cite{LondonSC+13,ChenSJ+15,ChenSJ+16}. We are now pursuing this kind of research in the 
academic arena and in the form of a technology start-up, NVision Imaging Technologies, that 
we have founded in 2015.

As should have become clear, much of my work has been influenced directly or indirectly by 
my postdoc time at Imperial College which played such an important role for my development 
as a scientist.

\section{Strolling through Quantum Optics -- Sanders}
\label{Sanders}
As one of Peter Knight's early postgraduate students, I have had the pleasure of a long and fruitful collaboration since I joined his group in 1986. I am honoured to contribute to this special issue a summary of my collaboration with Peter Knight. As a matter of fact, I also had the pleasure of working with Jon Marangos and Vlado Buzek on a special issue for Peter Knight's 60th birthday Festschrift, where our ``Foreword'' is (amusingly but somewhat appropriately) listed in online records as a ``Foreward''~\cite{SMB07}.

\subsection{Phase variables and squeezed states}
My doctoral work commenced with Peter Knight's prescient remark that squeezed light~\cite{PLKRodney} would soon be an experimental reality and that such states of light would have quantum noise that is phase sensitive, hence should be studied from the perspective of quantum phase operators. Together with Stephen Barnett, Peter Knight and I employed the Susskind-Glogower phase operator~\cite{SG64}
\begin{equation}
	\widehat{\text{e}^{\text{i}\phi}}
    	:=\left(\hat{n}+\mathds{1}\right)^{-1/2}      \hat{a}
        =\sum_{n=0}^\infty |n\rangle\langle n+1|,\;
   \widehat{\text{e}^{-\text{i}\phi}}
    	:=\hat{a}^\dagger\left(\hat{n}+\mathds{1}\right)^{-1/2}
        =\sum_{n=0}^\infty |n+1\rangle\langle n|
\end{equation}
for~$\hat{a}$ the harmonic-oscillator annihilation operator, to characterize the fluctuation properties of squeezed states and to construct number-phase uncertainty relations~\cite{SBK86}. The questions raised by our study provided an incentive to Stephen Barnett and David Pegg to discover their celebrated phase operator~\cite{PB88,PB89,PB89}.

\subsection{Quantum Walks}
Subsequent to my first foray into quantum-walk research, focusing on a theoretical study of quantum walks in higher dimensions~\cite{MBSS02}, Peter Knight expressed interest in realizing quantum walks as a quantum Galton board, or ``quantum quincunx'', which I learned from Peter Knight is a delightful alternative term for a Galton board. Travaglione and Milburn had suggested an insightful implementation in an ion trap~\cite{TM02}, and we proposed an alternative realization using cavity quantum electrodynamics~\cite{sanders02a}.

Our approach to quantum walks differed from that of Travaglione and Milburn's ion-trap implementation version. For~$\ket{n}$ a harmonic oscillator Fock number state and~$d$ some large dimensional cut-off, Travaglione and Milburn presumed that phase states~\cite{PB97}
\begin{equation}
	\left|\theta_k=2\pi\frac{k}{d}\right\rangle
    	=\frac{1}{\sqrt{d}}
        \sum_{n=0}^{d-1}
     \exp\left(\text{i}j\theta_k\right)
     \ket{n},
\end{equation}
for ionic harmonic motion could be prepared.
Instead we considered theoretically less-desirable but experimentally more reasonable (Glauber) coherent states~$\{\ket{\alpha};\alpha\in\mathbb{C}\}$ of the harmonic oscillator.

Rather than executing a quantum walk on a line, the concept is to execute a periodic sequence of two distinct unitary operations: one being a Hadamard gate that rotates the spin of the quantum coin from logical $\ket{0}$ and $\ket{1}$ states to an equally weighted superpositions of each and the other a rotation of the walker state around a circle in phase space according to the unitary transformation $\exp\left[(2\pi\text{i}/d)\hat{n}\otimes Z\right]$ for~$\hat{n}$ the harmonic-oscillator number operator and~$Z$ the Pauli operator leaving~$\ket{0}$ intact but changing the sign of~$\ket{1}$. Alternating between coin-flip operators and conditional rotations on the circle was challenging and later I worked with collaborators to devise a single-unitary-transformation approach relying on an indirect coin flip~\cite{XSBL08,XS08}

Peter Knight continued research on the quantum quincunx  through proposing an optical-cavity realization~\cite{knight03a}, and I am still active in this field including collaborating on photonic experiments~\cite{XQTS14,XZQ+15,BLQ+15}. The field of quantum walks is quite active due especially to applications in quantum algorithms, quantum transport and quantum simulation. Unfortunately the term ``quantum quincunx'' had less ``staying power'' than subject of quantum walks itself.

\subsection{Improving single-photon sources via linear optics and photodetection}
Motivated by the post-selected nonlinear-sign gate for linear-optical quantum computation~\cite{KLM01}, Peter Knight, I and others collaborated on improving triggered single-photon source efficiency, i.e., the probability for a single photon to appear in a pulse instead of vacuum, while avoiding multiphoton pulses. We established no-go results for enhancing efficiency via linear optics and conditional preparation based on photodetection~\cite{BSSK04,BSM+04}. Our work focused on reducing the general question of whether improving single-photon sources in this way could be reduced to a simple case from which all general results would follow. Eventually the simple case was solved, by Berry and Lvovsky in 2010~\cite{BL10}, where they showed that, without multiphoton components, the single-photon fraction cannot exceed the efficiency of the source under  this kind of linear optical processing. This tour-de-force settled the question we had been investigating.

Single-photon sources, and their improvement, continue to be of high importance in quantum optics, notably for BosonSampling~\cite{PLK,aaronson10a}. In fact the hardness of the BosonSampling problem relates to the counting-problem hardness of evaluating the permanent of the linear-optical unitary matrix to determining coincidence rate. We actually ran into the problem of the hardness of the permanent in our work on this single-photon efficiency enhancement~\cite{BSM+04}, which challenged our numerical simulations, but did not yet appreciate the importance of the permanent for computational complexity. Peter Knight continued work on single photons including his fine work concerning single photons on demand from tunable three-dimensional photonic band-gap structures~\cite{SFH+07}.
 
\subsection{Photon-number superselection and the entangled coherent-state representation}
Terry Rudolph and I initiated a debate on continuous-variable quantum teleportation ~\cite{RS01}, which led to a controversy~\cite{Wis04,Smo04} played an integral part in fostering what is now called quantum resource theories~\cite{BRS07}.
In response to claims that the coherent state offered simplicity, and by relying on entangled coherent states~\cite{sanders92,San12}, Peter Knight, Stephen Bartlett, Terry Rudolph and I co-developed an entangled-coherent-state representation, which provided an efficient, elegant description of quantum optical sources and detectors while respecting the photon-number superselection rule~\cite{SBRK03}.

The entangled-coherent state representation follows from the single-mode observation~\cite{BK91} that the number state can be expressed as a superposition of coherent states on the circle:
\begin{equation}
	\ket{n}
  	=\frac{1}{\sqrt{\Pi_n(m)}}
		\int_0^{2\pi}
	  	\frac{\text{d}\phi}{2\pi}
	      \text{e}^{-\text{i}n\phi}
			\ket{\sqrt{m}\text{e}^{\text{i}\phi}},\;
		\Pi_n(m)
	:=\text{e}^{-m}
          	\frac{m^n}{n!}
\end{equation}
for $\ket{\sqrt{m}\text{e}^{\text{i}\phi}}$ a coherent state with a mean of~$m$ photons.
The entangled coherent state version of this Fock-state expression with coherent states is given by
\begin{equation}
	\ket{n,n'}
    	=\frac{1}{\sqrt{\Pi_n(n)\Pi_{n'}(n')}}
        	\int_0^{2\pi}
           	\frac{\text{d}\phi}{2\pi}
           \int_0^{2\pi}
	          	\frac{\text{d}\phi'}{2\pi}
			     \text{e}^{-\text{i}(n\phi+n'\phi')}
	\ket{\sqrt{n}\text{e}^{\text{i}\phi},
    \sqrt{n'}\text{e}^{\text{i}\phi'}}
\end{equation}
which makes working with two-mode number state representations as simple as working with two-mode coherent-state representations.

The questions raised in our work, namely the requirement of coherence for claiming continuous-variable quantum teleportation~\cite{FSB+98}, became part of a bigger picture now referred to as quantum resource theory~\cite{BRS07,BG15}. In this framework, one considers a quantum resource required to overcome a superselection rule in order to complete a quantum information task.

\subsection{Complementarity and uncertainty relations for matter-wave interferometry}
My final work (so far!) with Peter Knight established a quantitative connection between interferometric duality  (which-way information vs fringe visibility) and Heisenberg's uncertainty relation for position and modular momentum. An apparent challenge arises to the Bohr-Heisenberg interpretation of the Heisenberg microscope due to employing a  metamaterial ``perfect lens'' to spontaneously emitted photons revealing which-path information. Specifically we show that complementarity is not compromised by superresolving detectors~\cite{MSK08}. \\

{\it Summary.--}
Peter Knight and I have collaborated on diverse topics from quantum optics to quantum information. Throughout I have benefited from Peter Knight's profound insight into problems, which stems from his unique talent of making the hardest problem simple and intuitive and thus tractable.

\section{Quantum engineering with hybrid quantum systems -- Twamley}
I was a Marie-Curie Research Fellow with Peter Knight and Tom Kibble from 1994-1995. Peter and I worked with his PhD student Joerg Steinbach to investigate the engineering of motional dynamics of trapped ions. However our discussions ranged more broadly over a host of topics - all informed by Peter's amazing recall for scientific results from the literature and his capacity for getting to the heart of a problem. I became particularly intrigued by the famous problem of the quantum mechanics of a light interacting with a mirror whose position must be treated in a quantum fashion. In 2014 Peter visited Macquarie University in Sydney for a month and with colleagues Keyu Xia and Mattias Johnasson we studied how to perform a quantum nondemolition measurement of single photons. We were able to use nonlinear optical processes and a fully multi-mode analysis to show how such measurements could be achieved. Peter's time at Macquarie not only helped our collaboration but students and other academics were lifted up by Peter's scientific effervescence and camaraderie.  

Engineering non-classical quantum states of motion of matter has long attracted much attention particularly if the object's motional properties admit the potential to generate macroscopic quantum superposition states. Early work on the preparation of non-classical motional states was theorised and performed in trapped ions  \cite{Cirac1993b, Meekhof1996a}, and  research studying the creation of non-classical motional states of individual ions and their uses was of interest in the late 90's  \cite{Gou1996g, Gou1996h, Bardroff1996b, Steinbach1997d}. Much of this work was performed in the so-called Lamb-Dicke (LD) regime where the photon recoil energy is much smaller than the motional quanta and thus the ion suffers little motional excitation. In \cite{Steinbach1997d}, we showed in detail how one can engineer the recoil dynamics of an ion trapped in two dimensions with anisotropic trap frequencies to laser engineer the motional Hamiltonian to be proportional to the generator of rotations between the $x-$ and $y-$ axes, or the $L_z$ component of angular momentum. We achieved this through the engineering of suitable Raman optical transitions so that, to leading order, the Hamiltonian did not entangle the motional and internal states of the ion. This work led to research later on which considered how one could use vibrational modes of motion for quantum computation \cite{paternostro2005}, and more recently in the imaging of single atoms in a fermionic quantum gas microscope \cite{haller2015}. 

More recently we have devised a similar protocol where a three level neutral atom is coupled to two optical cavities. The optical cavities are not directly coupled to each other but only  to the bridging atom. We showed that by controlling the electronic internal state of the atom one can enable a switchable optical coupling between the two optical cavities \cite{Xia2013}. Interestingly the switching is performed by optical manipulation of the atom and thus one has the potential for all-optical switching of both classical and quantum light signals. Following this we devised a scheme to implement an optical diode, where one direction of optical propagation is decreased in amplitude relative to the other using a three level atom with an unbalanced coupling from it's excited states to the two ground states without the use of a magnetic field \cite{Xia2014a}. Interesting this design was recently implemented and light diode operation was observed using a silicon nanowaveguide and either a weakly coupled  atomic ensemble  or strongly coupled single atom  \cite{Sayrin2015}. 

Optical photons make ideal hosts to store quantum information as they interact little with their environment. This latter effect however means that implementing quantum gates between individual photons is extremely challenging and this has led many to investigate the potential of linear optics to implement quantum computing. With the advent of strongly coupled atoms to optical nanofibres \cite{Kato2015}, there is real possibilities to achieve significantly large optical nonlinearities. However, since the late 90's researchers have studied the generation of quantum gates for single photons via optical nonlinearities and found many problems, with the result that although no proof existed that it was impossible - no successful protocol had been discovered. In 2016 however, in collaboration again with Peter Knight, we devised a scheme to implement a quantum non-demolition detection of a single photon which depended on the execution of a quantum gate between two photons \cite{Xia2016a}. We found that this could be achieved only by treating the problem in a full multimode fashion and carefully engineering the relative group velocities of the photons. Since this work another method to achieve quantum gates between single photons using counter propagating photons interacting with atoms has also been devised \cite{Combes2016}. 

The quantum engineering of the motion of objects is a rich and fertile topic. In \cite{Steinbach1997d}, we engineered a relatively simple dynamic: that of 2D rotation. More interesting would be to engineer non-classical motional states of large macroscopic objects e.g. Schrodinger cats. To achieve this would require a macroscopic motional system that is capable of possessing exceedingly large motional-$Q$ factors. Although initial hopes for this were proposed in optically trapped systems, the photon scattering noise present in optical tweezer traps has not provided a route to-date to achieve ultra-large motional-$Q$. In 2012 we proposed to achieve such long lived motional states using magnetically levitated systems \cite{Cirio2012}. This essentially uses the Meisner effect in superconductors to trap magnetic flux within a magnetic field. If the magnetic field is spatially inhomogeneous then any physical movement of the superconductor within this field would be resisted as this would entail a change in the threaded flux through the superconductor. This work showed the in-principle feasibility (in-theory), to trap and cool the motion of a large levitated object. More recently we developed this concept further to use the quantum magnetic fields produced by a superconducting flux qubit to force the magnetically levitated body into a macroscopic quantum spatial superposition \cite{Johnsson2016}. Using this additional leverage we showed how to implement a type of matter-wave interferometery which used the entire mass of the levitated object to perform absolute measurements of the local acceleration of gravity - or an absolute gravimeter. We estimated that the precision of such a device could be comparable or better than terrestrial atomic interferometers and would operate at a duty cycle of $\sim {\rm kHz}$, rather than Hz. The quest to engineer larger and larger Schrodinger cats brings one to the exciting question as to what happens if such spatial quantum superpositions involve active biological matter \cite{Toncang2017}, and whether such micro-Schrodinger cats are potentially experimentally achievable in the next few years. The field of hybrid quantum systems - where one quantum engineers various sub-systems together to achieve overall novel functionalities, is a promising new topic that holds much promise for the future.  

\section{A note on coherent states of light and their superpositions -- Vidiella-Barranco}
Coherent states of the oscillator were introduced by Schr\"odinger in the early days 
of quantum theory \cite{schroedinger26}. They were revived in the $1960$s within the context of quantum 
optics by Glauber \cite{glauber63}, following the development of laser radiation sources \cite{maiman60}.
It was later confirmed \cite{arecchi65} that a stabilized laser could generate in a straightforward way 
light having Poissonian photon statistics, i.e., coherent states, the ``quasi-classical", but quantum 
states of light (here denoted as $|\alpha\rangle$). Yet, their quantum superpositions (apart from a 
normalizing constant), e.g., $|\psi\rangle = |\alpha\rangle + \text{e}^{i\phi}|-\alpha\rangle$ 
may exhibit interesting non-classical features, such as squeezing \cite{janszky90,knight91,kennard27,PLKRodney}. 
Also, macroscopically distinguishable quantum 
superpositions of that type may be considered simplified, one-dimensional versions of Schr\"odinger cats 
\cite{schroedinger35}. During the $1990$s there was growing interest in such cat states, while I was working 
in my PhD at Imperial College under the supervision of Professor Knight. He suggested that I could investigate 
the properties of the cat states under dissipation. Some work regarding the influence of losses on quantum 
interference had been already done \cite{milburn85}, showing that the interference terms would decay much faster 
than the energy decay rate $\gamma$ (for $|\alpha|$ large enough). As a consequence a cat state would be rapidly 
reduced to a statistical mixture. Then I became involved in the investigation of the non-classical features of cats 
\cite{vidiella91} and also studied the influence of dissipation on them \cite{vidiella92a}. We found that those 
states may be sub-Poissonian, super-Poissonian or even Poissonian, and present squeezing (or not) depending on the 
relative phase $\phi$ between the component states \cite{vidiella91}. We also found that under dissipation, a 
property like (second-order) squeezing could survive longer than the interference terms themselves \cite{vidiella92a}. 
Surprisingly, higher-order squeezing may be momentarily generated during the dissipative evolution 
\cite{vidiella92a}. Also as a part of my PhD work, I investigated the interaction of cat states with atoms. 
In \cite{vidiella92b} it was suggested a way of discriminating cats from statistical mixtures of coherent states 
using the Jaynes-Cummings interaction \cite{jaynes63,knight93}. 

Meanwhile applications of quantum mechanics to information theory were on the rise \cite{bennett84,bennett93,shor94}. 
The quantum bits (qubits) are central elements in that framework, and interestingly, (continuous variable) 
coherent and cat states could also be used to implement logical qubits \cite{milburn99}, e.g. for quantum 
computation purposes \cite{kim02}. An important class of quantum superposition states involving two (or more) sub-systems 
are the strongly correlated states named entangled states \cite{horodecki09}. Known for a long time, 
\cite{einstein35}, they are nowadays considered important resources for quantum information.
In particular, one may define entangled coherent states for a bipartite system as \cite{sanders92,notation},
\begin{equation}
|\phi\rangle = |\alpha,-\alpha\rangle + |-\alpha,\alpha\rangle.
\end{equation}
Besides, generalizations of entangled states involving more sub-systems 
and based on coherent states were also envisaged. We could cite GHZ-type states \cite{an06},
\begin{equation}
|\varphi\rangle = |\alpha,\alpha,\alpha\rangle + |-\alpha,-\alpha,-\alpha\rangle, 
\end{equation}
as well as cluster-type entangled coherent states \cite{vidiella08,vidiella10},
\begin{equation}
|\xi\rangle =|\alpha,\alpha,\alpha,\alpha\rangle + |\alpha,\alpha,-\alpha-\alpha\rangle 
+ |-\alpha,-\alpha,\alpha,\alpha\rangle + |-\alpha,-\alpha,-\alpha,-\alpha\rangle.
\end{equation} 
Those continuous variables (entangled coherent) 
states may also be useful for tasks such as quantum computation or quantum teleportation 
\cite{milburn99,kim02,vidiella10}. Needless to say that the generation of cat states is of central importance
for some of those schemes, and we also contributed with proposals in trapped ions systems \cite{vidiella05a}
and in a modified Jaynes-Cummings model \cite{vidiella05b}.
Undoubtedly one of the most important applications of quantum mechanics in the 
information realm is the quantum key distribution (QKD) \cite{gisin02} method, for which coherent states of light 
play a prominent role. While protocols such as the BB84 \cite{bennett84} require single photon number states, 
most of the experiments actually employ strongly attenuated coherent states \cite{gisin02,gisin95}. 
However, protocols using brighter coherent states have also been proposed and executed \cite{grosshans02,grangier03}. 
I became involved again with coherent states, and proposed a QKD protocol using polarized coherent states 
\cite{vidiella06}, which is probably the first account employing thermal states as well. 
An advantage of using polarization encoding is that a separate local oscillator is not required. 
More recently, we presented an alternative QKD protocol using non-Gaussian, continuous variable states built from coherent states 
\cite{vidiella16}. The use of non-Gaussian states of light could be of relevance for the implementation of quantum repeaters 
in a QKD set up \cite{grangier11}. 

I clearly remember a nice and insightful lecture given by Professor Knight at Imperial College about coherent states while 
I was beginning my PhD, and I am grateful to him for showing me the captivating world of coherent states and their superpositions. 
\section{Remarks}
Peter Knight is one of the pioneers in quantum optics,
a missionary of the subject and a strong advocate
of the recent funding initiatives in quantum technologies. His research group at
Imperial College London was at the heart of theoretical quantum
optics and quantum information research for many decades. His leadership
has been pivotal for a coherent development of quantum
technologies in the UK. We all wish him a very happy 70th birthday. Thank you Peter.

\begin{acknowledgments}
SMB was supported by the Royal Society, grant no. RP150122.
 BMG acknowledges support from the UK EPSRC, grant  EP/M013294/1.
VK is supported by the UK EPSRC Grant EP/L022303/1. EP acknowledges the support of ``Research Projects for Excellence IKY/Siemens'' (Contract No. 23343).
BCS appreciates support from Alberta Innovates Technology Futures, NSERC and China's 1000 Talent Plan.

\end{acknowledgments}

\end{document}